\documentclass[epsfig,12pt]{article}
\usepackage{makeidx}
\usepackage{amsmath}
\usepackage{amsfonts}
\usepackage{amssymb}
\usepackage{graphicx}

\input epsf.sty
\newtheorem{theorem}{Theorem}
\newtheorem{acknowledgement}[theorem]{Acknowledgement}

\makeatletter \@addtoreset{equation}{section}
\renewcommand{\theequation}{\thesection.\arabic{equation}}
\input{tcilatex}
\begin{document}

\title{\vspace{-2cm}\rightline{\mbox{\small {LPHE-MS-15-09}} \vspace
{1cm}} \textbf{On Building superpotentials in F- GUTs}}
\author{E. H. Saidi\thanks{%
Email address: h-saidi@fsr.ac.ma} \\
{\small 1.}\emph{\ }{\small LPHE-Modeling and Simulations, Faculty Of
Sciences, Rabat, Morocco}\\
{\small 2. Centre of Physics and Mathematics, CPM- Morocco}\\
{\small 3- International Centre for Theoretical Physics, Miramare, Trieste,
Italy}}
\maketitle

\begin{abstract}
Using characters of finite group representations, we construct the fusion
algebras of operators of the spectrum of F- theory GUTs. These fusion
relations are used in building monodromy invariant superpotentials of the
low energy effective 4d $\mathcal{N}=1$ supersymmetric GUT models.\newline
\emph{Key words: F-GUT\ models, characters of finite groups, fusion algebra
of operators, superpotentials.}
\end{abstract}


\section{Introduction}

Ten dimensional superstring theory compactified to 4d space-time gives a
basic framework to describe elementary particle interactions in four
dimensions at M$_{GUT}$ scale. In this framework, one can build
phenomenologically viable supersymmetric Grand Unified Models (GUT) with 
\emph{discrete symmetries} covering results on minimal supersymmetric
standard model (MSSM) and aspects of neutrino physics \textrm{\cite{A}}
whose flavor mixing requires\textrm{\ }finite group symmetries\textrm{\ }%
like the alternating group $\mathbb{A}_{4}$ \textrm{\cite{B,C,4B,5B}}. One
also disposes of dual ways for engineering GUT models involving more
fundamental objects giving different, but equivalent, manners to approach
the idea of super-unification; for example by using heterotic string vacua
where both gravity and gauge dynamics descend from the closed string sector;
or by using type II strings with gauge degrees of freedom localised on
D-branes wrapping cycles of compact space. The string theory approach offers
therefore new tools beyond quantum field theory method to deal with the
usual difficulties in constructing GUT models \textrm{\cite{S}}. With the
new ingredients of 10d string theories compactified to 4d space-time; and
depending on the strength of the string couplings $g_{s}$, various proposals
have been developed to engineer stringy inspired 4d models extending MSSM;
and where discrete symmetries have a geometric interpretation and are
implemented in a natural way.

\  \  \  \  \newline
In the perturbative region of the string landscape where $g_{s}$ is small,
two particular approaches have been subject to intensive investigations;
these are: $\left( i\right) $ the approach based on $E_{8}\times E_{8}$
heterotic string taking advantage of the exceptional gauge symmetry to
realise the idea of grand unified theory with GUT symmetries of type $%
SU\left( 5\right) $ Georgi- Glashow symmetry, flipped $SU_{5}\times U\left(
1\right) $, $SO\left( 10\right) $ and $E_{6}$. These GUT symmetry candidates
are all of them subgroups of one of the two $E_{8}$'s of heterotic string
theory \textrm{\cite{1,2}}. $\left( ii\right) $ the approach based on
perturbative type IIA/B string orientifolds exploiting the localisation of
the gauge degrees of freedom along the D-branes \textrm{\cite{3,4,5}}.
Though it accommodates MSSM gauge group in a nice manner, the second
construction cannot implement exceptional gauge symmetry in terms of
D-branes. However, the difficulty to relate the $E_{8}$ of heterotic string
to type IIB D- brane construction can be overcome by thinking of $g_{s}$ as
a dynamical coupling that varies over the compactification space. In this
non perturbative regime, type IIB compactification with 7-branes is
described by F-theory \textrm{\cite{6}} where aspects of 7-branes\ get
geometrised in terms of properties of elliptic Calabi-Yau fourfolds (CY4)
with an $E_{8}$ geometry. This link is because at strong string coupling,
new degrees of freedom, namely the $\left( p,q\right) $- strings \textrm{%
\cite{61}}, become light and realise exceptional gauge symmetries even in a
theory based on branes; a special feature that makes F-theory on CY4s with
exceptional singularity a remarkable framework for the study of
supersymmetric GUT models building.

\  \  \  \  \newline
In the last few years, it has been shown that the set of four-dimensional
solutions of F- theory on elliptically Calabi-Yau fourfolds $\mathcal{Y}_{4}$
with exceptional $E_{8}$\ geometry constitutes a particularly interesting
class of string vacua for embedding supersymmetric GUTs in string theory.
The basic properties of the exceptional elliptic model singles out F-theory
GUT as the prototype where difficulties of intersecting brane models in
perturbative orientifolds are overcome. \textrm{More recently, there has bee}%
n an important development in embedding GUT-models with some special
discrete symmetries $\Gamma $ including the alternating $\mathbb{A}_{4}$
group privileged for neutrino flavor mixing \textrm{\cite{B,C}}. These kinds
of discrete groups $\Gamma $ emerge naturally in F-theory compactification
on elliptically Calabi-Yau 4-folds $\mathcal{Y}_{4}$ with a threefold base $%
\mathcal{B}_{3}$ and an $E_{8}$ geometry \textrm{\cite{1A}-\cite{13A}; }they
are captured by monodromy of 2-cycles in the compact sector of F-theory on
CY4. In this GUT building, one considers an exceptional 7-brane wrapping
divisors of $\mathcal{B}_{3}$\ and focus on the effective 8-dimensional
supersymmetric gauge theory on $\mathbb{R}^{1,3}\times \mathcal{S}_{{\tiny %
GUT}}$ combined with tools borrowed from heterotic spectral covers
construction \textrm{\cite{14A,14B,15A}}. The unified gauge theory $G_{%
{\tiny GUT}}\times \Gamma $ lives on wrapped 7-brane on GUT surface $%
\mathcal{S}_{{\tiny GUT}}$ with gauge symmetry $G_{{\tiny GUT}}$ controlled
by specific structure of the singularity over $\mathcal{S}_{GUT}$. This
singularity is given by the discriminant of the elliptic fibration and is
determined by the Kodaira's classification of singular fibers \textrm{\cite%
{15B,15C}}. The GUT gauge symmetry $G_{{\tiny GUT}}$ is engineered by
partial unfolding of the $E_{8}$\ singularity of the CY4 into $G_{{\tiny GUT}%
}\times U\left( 1\right) ^{n}$ with maximal abelian $U\left( 1\right)
^{n}\subset G_{\perp }$ and where $G_{\perp }$ is the commutant of $G_{%
{\tiny GUT}}$ in $E_{8}$. The discrete symmetry $\Gamma $ is given by
subgroups of the finite Weyl group of $G_{\perp }$. In the particular
example $G_{{\tiny GUT}}\times G_{\perp }=SO_{10}\times SU_{4}^{\bot }$,
matter curves of the GUT model are given by the decomposition of the $%
\mathbf{248}$ adjoint representation of $E_{8}$ in terms of $SO_{10}\times
SU_{4}^{\bot }$ representations namely%
\begin{equation}
\begin{tabular}{lll}
$\mathbf{248}$ & $\rightarrow $ & $\left( \mathbf{45},\mathbf{1}_{\perp
}\right) \oplus \left( \mathbf{1},\mathbf{15}_{\perp }\right) \oplus $ \\ 
&  & $\left( \mathbf{16},\mathbf{4}_{\perp }\right) \oplus \left( \overline{%
\mathbf{16}},\mathbf{\bar{4}}_{\perp }\right) \oplus \left( \mathbf{10},%
\mathbf{6}_{\perp }\right) $%
\end{tabular}
\label{a}
\end{equation}%
A similar decomposition can be also written down for the GUT\ models with
gauge symmetry $G_{{\tiny GUT}}=SU_{5}$ and commutant $G_{\perp
}=SU_{5}^{\bot }$. There, the $\mathbf{248}$ adjoint representation of $%
E_{8} $ is broken down into $SU_{5}\times SU_{5}^{\bot }$ representations;
and matter curves localised on brane intersections are associated with $%
\left( \mathbf{10},\mathbf{5}_{\perp }\right) ,$ $\left( \mathbf{\bar{5}},%
\mathbf{10}_{\perp }\right) $; their adjoints and $\left( \mathbf{1},\mathbf{%
24}_{\perp }\right) $.

\  \  \  \newline
In addition to gauge group $G_{{\tiny GUT}}$ representations, models of
F-theory GUTs have a finite spectrum $\left \{ \Phi _{R_{i}}\right \} $
indexed by quantum numbers of monodromy group $\Gamma $.\ In the example of $%
SO_{10}\times \Gamma $ models \textrm{\cite{1B}-\cite{8B}}, the possible
monodromies $\Gamma $ are given by subgroups of the permutation symmetry $%
\mathbb{S}_{4}$; and so have at most \emph{5} irreducible representations $%
\boldsymbol{R}_{i}$. For $SU_{5}\times \Gamma ^{\prime }$ models, the $%
\Gamma $'s are sub-symmetries of $\mathbb{S}_{5}$ having at most \emph{7}
irreducible representations. As shown on (\ref{a}), some of these $\Phi
_{R_{i}}$'s are somehow special in the sense they are scalars under the
gauge symmetry but carry non trivial charges under $\Gamma $. These special
fields representations, often called flavons, are also interesting in F-
theory GUT; in particular in the study of neutrino physics and in the
engineering of hierarchy \textrm{\cite{4B,5B}}. These flavons have been also
interpreted as extra fields of the Higgs sector like in extended MSSM; and
have been used for dealing with GUT constraints such as proton decay \textrm{%
\cite{9B,10B}. }By requiring invariance under $\Gamma $, one then disposes
of an important tool to construct chiral superfields $\Phi _{R_{i}}$
couplings including flavons; the $\Gamma $- invariance condition controls
therefore the structure of the superpotentials $W=W\left( \Phi
_{R_{i}}\right) $ of the supersymmetric GUT models since it will permit some
interactions between $\Phi _{R_{i}}$ and forbids others. However, to build
monodromy invariant superpotentials $W$ of the underlying low energy
effective 4d $\mathcal{N}=1$ supersymmetric QFT, one has to perform tensor
products $\otimes _{i}\Phi _{R_{i}}$ of representations $R_{i}$ of the
monodromy group $\Gamma $; and then takes the trace. These computations
require using fusion rules like 
\begin{equation}
\Phi _{R_{i}}\otimes \Phi _{R_{j}}=\dsum \limits_{R_{k}}\mathcal{C}%
_{R_{i},R_{j}}^{R_{k}}\Phi _{R_{k}}  \label{j}
\end{equation}%
which, to our knowledge, have not been enough studied in F- GUT literature 
\textrm{\cite{1C,4B}}. It is then interesting to explore this area and
determine these fusion rules and the corresponding closed algebras $\mathcal{%
F}_{\Gamma }$ for those discrete groups $\Gamma $ involved in F- theory
compactifications on CY4s.

\  \  \  \  \newline
In this paper, we derive the closed fusion algebras $\mathcal{F}_{\Gamma }$
of operators of the F- theory GUT spectrum $\left \{ \Phi _{R_{i}}\right \} $
by using algebraic properties of the $\Phi _{R_{i}}$'s; in particular the
characters $\mathbf{\chi }_{_{R_{i}}}$ of the group representations $R_{i}$
of monodromy $\Gamma $ and their dimensions. First, we show how these
operator fusion algebras $\mathcal{F}_{\Gamma }$ can be constructed; and as
applications, we give the explicit list of the $\mathcal{F}_{\Gamma }$'s for
those monodromy symmetries involved in the construction of superpotentials
in F-GUT; in particular for the cases of non abelian finite groups like the
symmetric groups $\mathbb{S}_{5},$ $\mathbb{S}_{4},$ $\mathbb{S}_{3};$ the
alternating $\mathbb{A}_{5},$ $\mathbb{A}_{4}$; and the dihedral $\mathbb{D}%
_{4}$. We also give the fusion algebras $\mathcal{F}_{\mathbb{Z}_{N}}$
associated with the particular abelian groups $\mathbb{Z}_{N}$ as a matter
to complete the study.

\  \  \  \newline
The presentation is as follows: In section 2, we give some useful tools and
properties on models of F-theory GUTs. First we recall the main lines of
F-theory on elliptic CY4s and the algebraic geometry approach using the Tate
form of the elliptic fibration. Then we describe the idea of spectral covers
construction in F-theory GUTs and show how it is used in practice for $%
SU_{5}\times \Gamma $ models and $SO_{10}\times \Gamma ^{\prime }$ models.
In section 3, we consider the example of $\mathbb{S}_{4}$ permutation group
and describe how this discrete symmetry appears as monodromy in F-GUT; and
how it is involved in building superpotentials $W\left( \Phi _{R_{i}}\right) 
$. Then, we use characters of the \emph{5} irreducible $\mathbb{S}_{4}$-
representations to build the underlying $\mathcal{F}_{\mathbb{S}_{4}}$
algebra of merging operators. In section 4, we derive the $\mathcal{F}%
_{\Gamma }$'s for the non abelian $\mathbb{A}_{4}$, $\mathbb{D}_{4}$ and $%
\mathbb{S}_{3}$ appearing also in the engineering of F-GUTs. In section 5,
we construct the $\mathcal{F}_{\Gamma }$'s for higher order groups; in
particular $\mathbb{S}_{5}$ and $\mathbb{A}_{5}$. In section 6, we conclude
and make a comment on the fusion algebra for abelian monodromies like $%
\mathbb{Z}_{N}$. In section 7, we give an appendix where useful tools on
discrete groups are collected.

\section{General on F- theory GUTs}

Since its discovery in 1996, F- theory \textrm{\cite{6}} and its
compactifications on elliptically fibered Calabi-Yau manifolds to lower
space-time dimensions have been subject to huge interest because of several
stringy and geometric properties; in particular for their dualities with
M-theory \textrm{\cite{1CA}} and heterotic string 
\begin{equation}
\begin{tabular}{lll}
F-theory/K3 & $\leftrightarrow $ & heterotic string /T$^{2}$%
\end{tabular}
\label{FH}
\end{equation}%
and also for those aspects linking brane physics with exceptional symmetry
groups to the homology of Calabi-Yau fourfolds with 4-form $G_{4}$ flux 
\textrm{\cite{1CB}}. The twelve dimensional F- theory compactified on
Calabi-Yau manifolds, which may be thought of as a non perturbative
description of a class of string vacua, can be motivated in various manners;
too particularly as a strongly coupled type IIB string theory with 7-branes
and varying dilaton. By using string dualities, it may be also linked to
M-theory on a vanishing 2-torus \textrm{\cite{1CC,1CD,1CE}}, 
\begin{equation}
T^{2}=S_{A}^{1}\times S_{B}^{1}
\end{equation}%
or remarkably to $E_{8}\times E_{8}$ heterotic string theory (\ref{FH})
where one disposes of basic results on engineering of vector bundles on
elliptically fibered Calabi-Yau 3-folds via the spectral covers construction 
\textrm{\cite{14A,14B,15A}}.\  \  \  \  \newline
In F-theory compactification, it is conjectured that physics of type IIB
orientifold \textrm{\cite{7,14B}} on complex $n$-fold $\mathcal{B}_{n}$ with
7-branes is encoded in the geometry of an $n+1$-fold $\mathcal{Y}_{n+1}$
given by a complex elliptic curve $\mathcal{E}$ fibered on the complex $%
\mathcal{B}_{n}$ base 
\begin{equation}
\begin{tabular}{lll}
$\mathcal{E}$ & $\rightarrow $ & $\mathcal{Y}_{n+1}$ \\ 
&  & $\downarrow $ \\ 
&  & $\mathcal{B}_{n}$%
\end{tabular}
\label{yn}
\end{equation}%
The curve fiber $\mathcal{E}$ is not part of the physical space-time; but a
clever trick that accounts for the variation of the complex structure $\tau $
of $\mathcal{E}\sim T^{2}$ with the two following features: $\left( i\right) 
$ the usual geometric $SL\left( 2,Z\right) $ action on the real 2- torus $%
T^{2}$ is identified with the well known $SL\left( 2,Z\right) $ symmetry of
10d type IIB supergravity supporting S- duality property. $\left( ii\right) $
the complex $\tau $ is realised in terms of complex axio- dilaton field like%
\begin{equation}
\tau =C+ie^{-\phi }
\end{equation}%
with axion $C$, dilaton $\phi $ and type IIB string coupling constant $%
e^{\phi }$. In this geometric representation, the location of the 7-branes
of IIB orientifold theory corresponds to singular value of the axio-dilaton $%
\tau $; which, from cycle- homology view, corresponds as well to the
shrinking of a 1-cycle of the elliptic fiber $\mathcal{E}$. Thus, the
degeneration locus of the curve $\mathcal{E}$ in F- theory on elliptically
fibered Calabi-Yau manifolds $\mathcal{Y}_{n+1}$ describes the presence of
7-branes wrapping cycles in the base $\mathcal{B}_{n}$. A simple example is
given by F-theory on complex K3 surface modeling physics of type IIB
orientifolds in $8d$ space- time dimensions. Another interesting example
corresponds to $4d$- space-time models given by F-theory on an elliptically
fibered Calabi-Yau fourfolds $\mathcal{Y}_{4}$ that captures the physics of
type IIB orientifolds on complex 3- folds $\mathcal{B}_{3}$.

\subsection{Tate models}

F-theory compactified on CY4s is modelled by using Weierstrass curve whose
useful features are nicely exhibited by using its Tate form. To fix idea, we
think it interesting to first review briefly some useful aspects on F-theory
and its compactification on elliptic CY4 hosting $\mathcal{N}=1$
supersymmetric GUTs; then turn to describe the main line lines of Tate
models.

\  \ 

$\bullet $ \emph{Weierstrass equation }\newline
Generally speaking, one of powerful features of F-theory is that it combines
two basic ingredients coming from two apparently different sources; one from
type II string and the other from heterotic string; these are:

\begin{description}
\item[$a)$] \  \ localisation of degrees of freedom as described in
perturbative models of type II orientifolds with D-branes,

\item[$b)$] \  \ exceptional gauge groups and spectral covers construction as
used in $E_{8}\times E_{8}$ heterotic string.
\end{description}

\  \  \  \  \newline
It happens that these two features are the essence of a mathematical theorem 
\textrm{\cite{8}} which states that every elliptic fibration like (\ref{yn})
can be represented by a Weierstrass model with an underlying exceptional
geometry. In the particular case of Calabi-Yau fourfolds $\mathcal{Y}_{4}$
based on threefolds $\mathcal{B}_{3}$, the corresponding elliptic fibration
with an E$_{8}$ geometry is described by the Weierstrass equation 
\begin{equation}
y^{2}=x^{3}+fxz^{4}+gz^{6}  \label{el}
\end{equation}%
with $\left( y,x,z\right) $ homogeneous coordinates of the weighted
projective space $WP_{2,3,1}$. The complex $f$ and $g$, specifying the shape
of the elliptic curve, are respectively holomorphic sections of $\mathbb{H}%
^{0}\left( \mathcal{B}_{3},\mathcal{K}^{-4}\right) $ and $\mathbb{H}%
^{0}\left( \mathcal{B}_{3},\mathcal{K}^{-6}\right) $ with $\mathcal{K}$ the 
\textrm{canonical bundle of the base} $\mathcal{B}_{3}$ \textrm{\cite{8A}}.
In the case where the base $\mathcal{B}_{3}$ is covered by some local
complex coordinates $\left \{ u_{i}\right \} $, the complex holomorphic
sections $f$ and $g$ are given by suitable polynomials%
\begin{equation}
f=f\left( u_{i}\right) \qquad ,\qquad g=g\left( u_{i}\right)
\end{equation}%
and moreover the elliptic fibration (\ref{el}) can be put into a Tate form
where the underlying $E_{8}$ gauge symmetry is broken down to some gauge
group $G_{_{GUT}}$ along the complex surface divisor $\mathcal{S}_{_{GUT}}$
of the base space $\mathcal{B}_{3}$ of the fibration; see eq(\ref{tat})
reported below. Notice also that in the coordinates patch where\ $z=1$; eq(%
\ref{el}) reduces to 
\begin{equation}
y^{2}=x^{3}+fx+g
\end{equation}%
with discriminant $\Delta $ given by the usual formula 
\begin{equation}
\Delta =27g^{2}+4f^{3}  \label{di}
\end{equation}%
The zero values of this discriminant $\Delta $ play an important role in the
F-GUT construction. By thinking of $\Delta $ in terms of a product of
factors like $\dprod \nolimits_{i}\Delta _{i}$, its zeros describe the loci $%
\Delta _{i}=0$ where the elliptic curve $\mathcal{E}$ degenerates. These
loci are interpreted in terms of locations of the 7-branes wrapping some
divisors $\mathcal{D}_{i}$ of the $\mathcal{B}_{3}$- base of the CY4. The
extra non compact directions of the 7-branes fill the \emph{4d} space- time
where live GUT models.\ To get more insight into these brane/geometry
features and on the way the gauge symmetry groups and matter localisations
emerge in F-theory description, it is interesting to use the Tate form of
the elliptic curve fiber $\mathcal{E}$ that we describe in what follows.

\  \  \  \  \  \ 

$\bullet $ \emph{Tate representation }\newline
A convenient way to exhibit explicitly the singularities of the elliptic
fibration (\ref{el}) is to use the Tate form of the elliptic curve; it is
given by the following complex holomorphic equation \textrm{\cite{9} } 
\begin{equation}
y^{2}=x^{3}+a_{1}xyz+a_{2}x^{2}z^{2}+a_{3}yz^{3}+a_{4}xz^{4}+a_{6}z^{6}
\label{tat}
\end{equation}%
which is \textrm{related to Weierstrass eq(\ref{el}) by} coordinates
redefinition. Like for the complex $f$ and $g$, the new complex holomorphic
sections $a_{n}=a_{n}\left( u_{i}\right) $ depend on the complex coordinates 
$u_{i}$ of base $\mathcal{B}_{3}$; they encode properties of the
discriminant loci $\Delta _{i}=0$ of the above elliptic fibration obtained
by solving the condition%
\begin{equation}
\Delta =-\frac{1}{4}\beta _{2}^{2}\left( \beta _{2}\beta _{6}-\beta
_{4}^{2}\right) -8\beta _{4}^{3}-27\beta _{6}^{2}+9\beta _{2}\beta _{6}\beta
_{4}  \label{id}
\end{equation}%
where we have set%
\begin{equation}
\begin{tabular}{lll}
$\beta _{2}$ & $=$ & $a_{1}^{2}+4a_{2}$ \\ 
$\beta _{4}$ & $=$ & $a_{1}a_{3}+2a_{4}$ \\ 
$\beta _{6}$ & $=$ & $a_{3}^{2}+4a_{6}$%
\end{tabular}
\label{dd}
\end{equation}%
Notice that $f$ and $g$ of (\ref{di}) are related to the $a_{n}$'s like%
\begin{equation}
\begin{tabular}{lll}
$f$ & $=$ & $\frac{1}{24}\left( \beta _{2}^{2}-24\beta _{4}\right) $ \\ 
$g$ & $=$ & $-\frac{1}{864}\left( 36\beta _{2}^{2}\beta _{4}-\beta
_{3}^{2}-216\beta _{6}\right) $%
\end{tabular}%
\end{equation}%
As noticed before the discriminant (\ref{id}-\ref{dd}) can, under some
assumptions\textrm{\footnote{%
Near GUT surface $\mathcal{S}_{GUT}$ of the $SU_{5}$ model defined by the
divisor $w=0$, the $a_{n}$ holomorphic sections of $\mathcal{B}_{3}$ have
the typical form $a_{n}\sim w^{5-k}\boldsymbol{b}_{k,n}$ where the new $%
\boldsymbol{b}_{k,n}=\boldsymbol{b}_{k,n}\left( u_{1},u_{2},w\right) $ are
as in table (\ref{tab}).}} on the $a_{n}$'s, be factorized into product of
factors like $\dprod \nolimits_{i}\Delta _{i}$; each factor $\Delta _{i}$
describing the location of a 7- brane on a divisor $\mathcal{D}_{i}$ in the
complex $3d$ base $\mathcal{B}_{3}$; one of them is the GUT surface $%
\mathcal{S}_{_{GUT}}$. Two divisors $\mathcal{D}_{i},$ $\mathcal{D}_{j}$ may
intersect on curves $\Sigma _{ij}=\mathcal{D}_{i}\cap \mathcal{D}_{j}$ where
fundamental matter localise; while three divisors may intersect at points $%
P_{ijk}=\mathcal{D}_{i}\cap \mathcal{D}_{j}\cap \mathcal{D}_{k}$
corresponding to Yukawa couplings. It turns out that the gauge symmetry
group on GUT surface $\mathcal{S}_{_{GUT}}$ is precisely encoded by the
vanishing degree of the $a_{n}$ sections on $\mathcal{S}_{_{GUT}}$\textrm{. }%
For the examples of $SU\left( 5\right) $ and $SO\left( 10\right) $ gauge
symmetries along GUT divisor $\mathcal{S}_{GUT}$ given by $w=0$; we have%
\textrm{\ }the following behaviors,; for more details see \textrm{\cite%
{9,10,11}}%
\begin{equation}
\begin{tabular}{l|lllll}
group & $a_{1}$ \  \  \  \  \  \  \  & $a_{2}$ \  \  \  \  \  \  \  & $a_{3}$ \  \  \  \  \
\  \  & $a_{4}$ \  \  \  \  \  \  \  & $a_{6}$ \\ \hline
$SU\left( 5\right) $ & $\boldsymbol{b}_{5}$ & $\boldsymbol{b}_{4}w$ & $%
\boldsymbol{b}_{3}w^{2}$ & $\boldsymbol{b}_{2}w^{3}$ & $\boldsymbol{b}%
_{0}w^{5}$ \\ 
$SO\left( 10\right) $ & $\boldsymbol{b}_{5}w$ & $\boldsymbol{b}_{4}w$ & $%
\boldsymbol{b}_{3}w^{2}$ & $\boldsymbol{b}_{2}w^{3}$ & $\boldsymbol{b}%
_{0}w^{5}$%
\end{tabular}
\label{tab}
\end{equation}%
where the complex $\boldsymbol{b}_{k}$'s generically depend on all
coordinates of $\mathcal{B}_{3}$ but do not contain an overall factor of the
complex variable $w$.\textrm{\ }Using the expression of the $a_{k}$'s in
terms of $\boldsymbol{b}_{k}$ and $w$, the discriminant $\Delta _{su_{5}}$
of the elliptic fibration with $SU_{5}$ symmetry on GUT surface $\mathcal{S}%
_{_{GUT}}$ factorises as%
\begin{equation}
\Delta _{su_{5}}=-w^{5}\times \delta  \label{w5}
\end{equation}%
with $w^{5}$ encoding $SU_{5}$ symmetry and the factor $\delta $ describing
a single- component locus of an $I_{1}$ singularity of Kodaira
classification; it reads as follows%
\begin{equation}
\delta =\boldsymbol{b}_{5}^{4}\mathcal{P}+w\boldsymbol{b}_{5}^{2}\left(
8b_{4}\mathcal{P}+b_{5}\mathcal{R}\right) +w^{2}\left( 16\boldsymbol{b}%
_{3}^{2}\boldsymbol{b}_{4}^{2}+\boldsymbol{b}_{5}\mathcal{Q}\right) +%
\mathcal{O}\left( w^{3}\right)
\end{equation}%
where%
\begin{equation}
\begin{tabular}{lll}
$\mathcal{P}$ & $=$ & $\boldsymbol{b}_{3}^{2}\boldsymbol{b}_{4}-\boldsymbol{b%
}_{2}\boldsymbol{b}_{3}\boldsymbol{b}_{5}+\boldsymbol{b}_{0}\boldsymbol{b}%
_{5}^{2}$ \\ 
$\mathcal{R}$ & $=$ & $4\boldsymbol{b}_{0}\boldsymbol{b}_{4}\boldsymbol{b}%
_{5}-\boldsymbol{b}_{3}^{3}-\boldsymbol{b}_{2}^{2}\boldsymbol{b}_{5}$%
\end{tabular}%
\end{equation}%
From the factorisation (\ref{w5}), we learn that the cohomology class $\left[
\Delta \right] $ of the discriminant $\Delta _{su_{5}}$ in terms of the GUT
divisor $\left[ S\right] $ and the $I_{1}$ singularity divisor $\left[ S_{1}%
\right] $ is given by the sum 
\begin{equation}
\left[ \Delta \right] =5\left[ S\right] +\left[ S_{1}\right]  \label{5w}
\end{equation}

\subsection{Spectral covers in $SO_{10}$ and $SU_{5}$ models}

In this subsection, we first describe the main lines of spectral covers in
F- theory on Calabi-Yau fourfolds. Then, we focus on how the construction
works on the example of GUT- models embedded in F-theory compactifications.
We restrict this description to those $G\times \Gamma $ models with discrete 
$\Gamma $ and gauge invariance $G=SU_{5},$ $SO_{10}$.

\subsubsection{Spectral covers in F- theory}

In F- theory based models building, spectral covers approach provides a
tricky manner to: $\left( i\right) $ determine the various matter
representations $\mathcal{R}_{i}$ localised along the matter curves $\Sigma
_{i}$ as exhibited by eq(\ref{eb}) given below; and $\left( ii\right) $
engineer the flux required by chirality feature of the model in terms of few
parameters; see eqs(\ref{bb}-\ref{be}). If one is interested only in the
physics on the GUT surface $\mathcal{S}_{_{GUT}}$, the key idea of the
spectral covers construction proceeds as follows:

\begin{itemize}
\item first, use Tate form (\ref{tat}) with holomorphic sections $a_{n}$ of
table (\ref{tab}) to fix the desired gauge symmetry $G$ on GUT surface $%
\mathcal{S}_{_{GUT}}$. For the case where $G=SU\left( 5\right) $, we have%
\begin{equation}
y^{2}=x^{3}+b_{5}xy+b_{4}x^{2}w+b_{3}yw^{2}+b_{2}xw^{3}+b_{0}w^{5}
\end{equation}%
In this case the initial $E_{8}$ singularity of the elliptic fibration of
the CY4 has been lifted to $G=SU\left( 5\right) \times U\left( 1\right) ^{4}$
with $U\left( 1\right) ^{4}$ standing for the Cartan charges of the
perpendicular $SU\left( 5\right) ^{\bot }$, the commutant of the $SU\left(
5\right) $ gauge symmetry inside $E_{8}$.

\item second, restrict the Tate model to the neighbourhood of the divisor $%
\mathcal{S}_{_{GUT}}\subset \mathcal{B}_{3}$ by using spectral covers
method. The latter is inspired from spectral covers construction used in
building models embedded in heterotic string theory \textrm{\cite{14B}}. In
F-theory, the idea of the spectral covers method relies on zooming into the
local neighbourhood of the $w=0$ divisor $\mathcal{S}_{_{GUT}}$ inside $%
\mathcal{B}_{3}$ by dropping out all terms of higher power in the normal
coordinate $w$ that appear in the sections $\boldsymbol{b}_{n}$; that is
restricting them to 
\begin{equation}
b_{n}=\left. \boldsymbol{b}_{n}\right \vert _{w=0}
\end{equation}%
where now $b_{n}$ live on $\mathcal{S}_{_{GUT}}$.

\item then, think of $\mathcal{S}_{_{GUT}}$ as the base of the bundle $%
\mathcal{K}_{S}\rightarrow \mathcal{S}_{_{GUT}}$, with GUT surface given by $%
s=0$; and approach the neighbourhood of $\mathcal{S}_{_{GUT}}$ in terms of
spectral surfaces $\mathcal{C}_{n}$ given by divisors of the total space. In
case of SU$_{5}$ model, the integer n takes the values $5,$ $10,$ $20$
respectively associated with 10-plets 10$_{t_{i}}$, 5-plets 5$_{t_{i}+t_{j}}$
and charged flavons $\vartheta _{t_{i}-t_{j}}$. For the example of $\mathcal{%
C}_{5}$, describing the fundamental representation of $SU\left( 5\right)
^{\bot }$, we have%
\begin{equation}
\mathcal{C}_{5}=b_{0}s^{5}+b_{1}s^{4}+b_{2}s^{3}+b_{3}s^{2}+b_{4}s+b_{5}=0
\label{c5}
\end{equation}%
with $b_{1}=0$ required by traceless property of $SU\left( 5\right) $; a
feature that is implemented explicitly by factorising $\mathcal{C}_{5}$ as
follows%
\begin{equation}
\mathcal{C}_{5}=b_{0}\dprod \limits_{i=1}^{5}\left( s-t_{i}\right)
\label{5c}
\end{equation}%
and requiring 
\begin{equation}
t_{1}+t_{2}+t_{3}+t_{4}+t_{5}=0.
\end{equation}%
One can think about above $\mathcal{C}_{5}$, whose expression (\ref{5c}) is
manifestly invariant under $\mathbb{S}_{5}$ permuting the 5 roots t$_{i}$,
as encoding information about the discriminant locus in the local vicinity
of GUT surface.

\item finally, fluxes are engineered by splitting spectral covers $\mathcal{C%
}_{n}$ like $\dprod \nolimits_{k}\mathcal{C}_{n_{k}}$ with $n=\sum_{k}n_{k}$
and where each factor $\mathcal{C}_{n_{k}}$ has an expansion in terms of the
spectral variable s as in eq(\ref{c5}). This splitting introduce new
holomorphic sections obeying constraints following by equating the expansion
of $\mathcal{C}_{n}$ with the one resulting from $\dprod \nolimits_{k}%
\mathcal{C}_{n_{k}}$. As an example, the splittings of $\mathcal{C}_{5}$ and
corresponding monodromy groups are given by%
\begin{equation}
\begin{tabular}{l|ll}
splitted spectral covers &  & \  \  \  \  \  \  \  \ monodromy \  \  \  \  \  \  \  \  \ 
\\ \hline
$\  \  \mathcal{C}_{4}^{\left( 5\right) }\times \mathcal{C}_{1}^{\left(
5\right) }$ &  & \  \  \  \  \  \  \  \  \ $\  \mathbb{S}_{4}$ \\ 
$\  \  \mathcal{C}_{3}^{\left( 5\right) }\times \mathcal{C}_{2}^{\left(
5\right) }$ &  & \  \  \  \  \  \  \  \  \ $\mathbb{S}_{3}\times $ $\mathbb{S}_{2}$
\\ 
$\  \  \mathcal{C}_{3}^{\left( 5\right) }\times \mathcal{C}_{1}^{\left(
5\right) }\times \mathcal{C}_{1}^{\left( 5\right) }$ &  & \  \  \  \  \  \  \  \  \ $%
\mathbb{S}_{3}$ \\ 
$\  \  \mathcal{C}_{2}^{\left( 5\right) }\times \mathcal{C}_{2}^{\left(
5\right) }\times \mathcal{C}_{1}^{\left( 5\right) }$ &  & \  \  \  \  \  \  \  \  \ $%
\mathbb{S}_{2}\times $ $\mathbb{S}_{2}$ \\ 
$\  \  \mathcal{C}_{2}^{\left( 5\right) }\times \mathcal{C}_{1}^{\left(
5\right) }\times \mathcal{C}_{1}^{\left( 5\right) }\times \mathcal{C}%
_{1}^{\left( 5\right) }$ &  & \  \  \  \  \  \  \  \  \ $\mathbb{S}_{2}$ \\ 
$\  \  \mathcal{C}_{1}^{\left( 5\right) }\times \mathcal{C}_{1}^{\left(
5\right) }\times \mathcal{C}_{1}^{\left( 5\right) }\times \mathcal{C}%
_{1}^{\left( 5\right) }\times \mathcal{C}_{1}^{\left( 5\right) }$ &  & \  \  \
\  \  \  \  \  \ - \\ \hline
\end{tabular}
\label{bb}
\end{equation}%
where for instance%
\begin{equation}
\begin{tabular}{lll}
$\mathcal{C}_{4}^{\left( 5\right) }$ & $=$ & $\alpha _{0}s^{4}+\alpha
_{1}s^{3}+\alpha _{2}s^{2}+\alpha _{3}s+\alpha _{4}$ \\ 
$\mathcal{C}_{1}^{\left( 5\right) }$ & $=$ & $\beta _{0}s+\beta _{1}$%
\end{tabular}%
\end{equation}%
and similar relations for the others. The $\alpha _{l}$'s and $\beta _{l}$'s
are new holomorphic sections; they are related to the $b_{l}$'s like%
\begin{equation}
\begin{tabular}{lllll}
$b_{0}=\alpha _{0}\beta _{0}$ &  & , &  & $b_{1}=\alpha _{0}\beta
_{1}+\alpha _{1}\beta _{0}$ \\ 
$b_{2}=\alpha _{1}\beta _{1}+\alpha _{2}\beta _{0}$ &  & , &  & $%
b_{3}=\alpha _{2}\beta _{1}+\alpha _{3}\beta _{0}$ \\ 
$b_{4}=\alpha _{3}\beta _{1}+\alpha _{4}\beta _{0}$ &  & , &  & $%
b_{5}=\alpha _{4}\beta _{1}$%
\end{tabular}
\label{be}
\end{equation}
\end{itemize}

\  \emph{Localised matter}\  \  \  \newline
In the spectral covers description introduced just above, the various matter
representations $\mathcal{R}_{n}$ localised along the matter curves $\Sigma
_{n}$ are determined by the intersections $\mathcal{C}_{n}\cap \mathcal{S}%
_{_{GUT}}$. For the example of the fundamental $\mathcal{C}_{5}$, the
intersection with the GUT surface $\mathcal{S}_{_{GUT}}$ is given by the
relation 
\begin{equation}
b_{5}=-b_{0}t_{1}t_{2}t_{3}t_{4}t_{5}=0
\end{equation}%
having five solutions given by $t_{i}=0$. These solutions describe precisely
the localisation of five tenplet matter curves of the $SU\left( 5\right)
\times \mathbb{S}_{5}$ model 
\begin{equation}
10_{t_{1}},10_{t_{2}},10_{t_{3}},10_{t_{4}},10_{t_{5}}  \label{eb}
\end{equation}%
By extending this spectral covers construction to the other representations
of $SU\left( 5\right) ^{\bot }$ involved in the breaking of E$_{8}$ down to $%
SU\left( 5\right) \times SU\left( 5\right) ^{\bot }$, in particular to the
antisymmetric and adjoint ones, one can also describe in quite similar
manner the localisation of the other matter multiplets namely the $\left( 
\mathbf{\bar{5}},\mathbf{10}_{\perp }\right) $ and $\left( \mathbf{1},%
\mathbf{24}_{\perp }\right) $; for explicit details see \textrm{\cite{2B}}
and refs therein.

\  \  \ 

\emph{heterotic dual}\newline
First notice that not any F-theory compactification has a heterotic string
dual. For those cases of F-theory compactifications having heterotic string
duals; the elliptic Calabi-Yau fourfolds $\mathcal{Y}_{4}:\mathcal{E}%
\rightarrow \mathcal{B}_{3}$ have also a K3-fibration over a complex surface 
$\mathcal{B}_{2}$ as follows 
\begin{equation}
\mathcal{Y}_{4}:K3\rightarrow \mathcal{B}_{2}
\end{equation}%
By thinking of the complex surface $K3$ in term of the elliptic fibration of
a real 2-torus T$^{2}$ over a real 2-sphere S$^{2}$; or more precisely in
terms of a complex elliptic curve over projective line like $K3$ $:\mathcal{E%
}\rightarrow \mathbb{P}^{1}$; it follows that the complex 3d base space is
in turns given by a fibration of a complex projective line $\mathbb{P}^{1}$
on complex base surface $\mathcal{B}_{2}$ as follows 
\begin{equation}
\mathcal{B}_{3}:\mathbb{P}^{1}\rightarrow \mathcal{B}_{2}  \label{3b}
\end{equation}%
This fibration of $\mathcal{B}_{3}$ puts therefore a strong restriction on
the set of Calabi-Yau fourfolds of F- theory GUT models having heterotic
string duals.

\  \ 

\emph{ALE fibration}\  \  \  \newline
The fibration (\ref{3b}) is very suggestive in dealing with local models of
F-theory-GUT. There, one has a quite similar local structure of $\mathcal{B}%
_{3}$ near $\mathcal{S}_{_{GUT}}$ since the role of $\mathcal{B}_{2}$ is
done but $\mathcal{S}_{_{GUT}}$; and the role of $\mathbb{P}^{1}$ in (\ref%
{3b}) gets now played by several intersecting $\mathbb{P}_{i}^{1}$'s glued
as in the graph of Dynkin diagram of Lie algebra $E_{8}$. In the limit where
all sizes of the $\mathbb{P}_{i}^{1}$'s are shrunk to zero, one is left with
an E$_{8}$ singularity of the elliptic fibration.%
\begin{equation}
\mathcal{B}_{3}:\left. ALE\right \vert _{E_{8}}\rightarrow \mathcal{S}%
_{_{GUT}}
\end{equation}
By blowing up the size of some of the $\mathbb{P}_{i}^{1}$'s, one can
engineer desired gauge symmetries $G$ given by subgroups of $E_{8}$.
Therefore, the complex surface $\mathcal{S}_{_{GUT}}$ can be locally viewed
as the basis of an ALE fibration which describes the singularity structure
along $\mathcal{S}_{_{GUT}}$. The ALE fiber contains a distinguished set of
two-cycles $\gamma _{i}$ with intersection $\gamma _{i}\circ \gamma _{j}$
given by minus the Cartan matrix of $E_{8}$. 
\begin{equation}
\gamma _{i}\circ \gamma _{j}=-K_{ij}\left( E_{8}\right)
\end{equation}%
If a number r of two-cycles $\gamma _{i}^{\prime }$ among the eight ones
have non-zero size; the $E_{8}$ symmetry is broken to subgroups $%
G_{8-r}\times U\left( 1\right) ^{r}$; for example where $r=4$, the $E_{8}$
symmetry breaks down to $SU_{5}\times U\left( 1\right) ^{4}$; and for $r=3$
it breaks to $SO_{10}\times U\left( 1\right) ^{3}$. Moreover, matter curves
and Yukawa coupling points on the divisor $\mathcal{S}_{_{GUT}}$ exhibit
enhanced gauge symmetries; they correspond to brane- intersections where
localise matter and Yukawa interactions. In what follows, we consider the
cases of $G\times \Gamma $ models with $G=SO_{10},$ $SU_{5}$; and comment
briefly on the spectral cover construction for some discrete monodromies $%
\Gamma $.

\subsubsection{$G\times \Gamma $ models: $G=SO_{10},$ $SU_{5}$}

\begin{figure}[tbph]
\begin{center}
\hspace{0cm} \includegraphics[width=6cm]{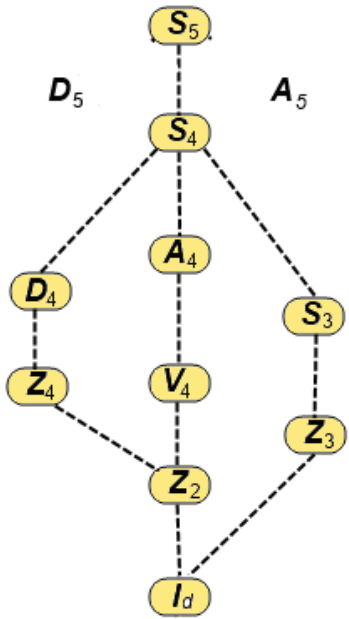}
\end{center}
\par
\vspace{-0.5 cm}
\caption{A branch of the tree of the $156$ subgroups of the symmetric $S_{5}$%
. The branch with top vertex $S_{4}$, gives the \emph{30} subgroups of the
symmetric $S_{4}\subset S_{5}$.}
\label{5}
\end{figure}

Focusing first on the family $SO_{10}\times \Gamma $ models of F-theory GUTs
with $SO_{10}$ gauge symmetry, the candidates for discrete monodromy $\Gamma 
$ is given by one of the \emph{30} possible subgroups of the symmetric group 
$\mathbb{S}_{4}$; the Weyl group group of $SU_{4}^{\bot }$. To make an idea
on the explicit list of these monodromy $\Gamma $'s, see eq(\ref{ls})
reported below and also the $\mathbb{S}_{4}$- branch in fig.\ref{5} giving
particular subgroups of $\mathbb{S}_{5}$.

\  \  \  \ 

$\bullet $ $SO_{10}\times \Gamma $ models\newline
The matter content $\left \{ \Phi _{R_{i}}\right \} $ of the $SO_{10}\times
\Gamma $ models are read from the decomposition (\ref{a}); it is labeled by
four weights t$_{i}$ like%
\begin{equation}
\begin{tabular}{llllll}
$\Phi _{R_{i}}$ & : & $\mathbf{16}_{t_{i}},$ & $\mathbf{16}_{-t_{i}},$ & $%
\mathbf{10}_{t_{i}+t_{j}},$ & $\mathbf{1}_{t_{i}-t_{j}}$%
\end{tabular}
\label{b}
\end{equation}%
with traceless condition 
\begin{equation}
t_{1}+t_{2}+t_{3}+t_{4}=0  \label{c}
\end{equation}%
The components of the four sixteen-plets $\mathbf{16}_{t_{i}}\equiv \left \{ 
\mathbf{16}_{t_{1}},\mathbf{16}_{t_{2}},\mathbf{16}_{t_{3}},\mathbf{16}%
_{t_{4}},\right \} $ and those of the six ten-plets $\mathbf{10}%
_{t_{i}+t_{j}}\equiv \left \{ \mathbf{10}_{\pm \left( t_{1}+t_{2}\right) },%
\mathbf{10}_{\pm \left( t_{1}+t_{3}\right) },\mathbf{10}_{\pm \left(
t_{2}+t_{3}\right) }\right \} $ as well as the \emph{15} singlets (flavons)
are related to each other by monodromies $\Gamma $. These discrete
symmetries offer a framework of approaching $SO_{10}\times \Gamma $ models
embedded in F-theory compactified on elliptic Calabi-Yau fourfolds%
\begin{equation}
CY4\sim E_{{\tiny SO}_{{\tiny 10}}}\times \mathcal{B}_{3}
\end{equation}%
with complex 3- $\dim $ base $\mathcal{B}_{3}$ containing $\mathcal{S}_{GUT}$%
. The Tate form of this Calabi-Yau fourfolds is realised as follows%
\begin{equation}
y^{2}=x^{3}+b_{5}xyw+b_{4}x^{2}w+b_{3}yw^{2}+b_{2}xw^{3}+b_{0}w^{5}
\end{equation}%
with holomorphic sections $b_{k}$ living on the GUT surface. The homology
classes of $x,$ $y,$ $w$ and $b_{k}$ are expressed in terms of the Chern
class $c_{1}=c_{_{1}}\left( \mathcal{S}_{GUT}\right) $ of the tangent bundle
of the $\mathcal{S}_{GUT}$ surface; and the Chern class $-t$ of the normal
bundle $\mathcal{N}_{\mathcal{S}_{GUT}|\mathcal{B}_{3}}$ as follows%
\begin{equation}
\begin{tabular}{lll}
$\left[ y\right] $ & $=$ & $3\left( c_{1}-t\right) $ \\ 
$\left[ x\right] $ & $=$ & $2\left( c_{1}-t\right) $ \\ 
$\left[ w\right] $ & $=$ & $-t$ \\ 
$\left[ b_{k}\right] $ & $=$ & $\left( 6c_{1}-t\right) -kc_{1}$%
\end{tabular}
\label{31}
\end{equation}%
Matter curves in $SO_{10}\times \Gamma $ models are described by spectral
covers of GUT surfaces. To each of the multiplets in (\ref{b}); it is
associated a spectral cover $\mathcal{C}_{n}$ given by an order n
holomorphic polynomial in a spectral variable $s$; with number of roots
given by dimension of corresponding $SU_{4}^{\bot }$ representation. For
example, the spectral cover $\mathcal{C}_{4}$ associated with the four
sixteen- plets $\mathbf{16}_{t_{i}}$ is given by 
\begin{equation}
\mathcal{C}_{4}:b_{0}s^{4}+b_{1}s^{3}+b_{2}s^{2}+b_{3}s+b_{4}=0
\end{equation}%
with $b_{1}=0$ due to traceless condition of $SU_{4}^{\bot }$. This
polynomial factorises like%
\begin{equation}
\mathcal{C}_{4}=b_{0}\dprod \limits_{i=1}^{4}\left( s-t_{i}\right)
\end{equation}%
where the $t_{i}$ zeros are precisely as in eq(\ref{c}). The last expression
of $\mathcal{C}_{4}$ is manifestly invariant under $\mathbb{S}_{4}$
permuting the 4 roots t$_{i}$,. Matter curves $\mathbf{16}_{t_{i}}$ are
given by the intersection of $\mathcal{C}_{4}$ with the GUT surface $%
\mathcal{S}_{GUT}$ realised in this formulation by the divisor $s=0$; that is%
\begin{equation}
\mathcal{C}_{4}\cap \mathcal{S}_{GUT}\qquad \Rightarrow \qquad b_{4}=0\qquad
\Rightarrow \qquad b_{0}\dprod \limits_{i=1}^{4}t_{i}=0
\end{equation}%
Similar expression can we written down for the other spectral covers; for
the example of the six tenplets $\mathbf{10}_{t_{i}+t_{j}}$; the
corresponding spectral cover is given by 
\begin{equation}
\mathcal{C}_{6}=\dsum \limits_{k=1}^{6}d_{k}s^{6-k}
\end{equation}%
with the six matter curves $\mathbf{10}_{t_{i}+t_{j}}$ on GUT surface
localised at the zeros of $\mathcal{C}_{6}$ as shown below%
\begin{equation}
\mathcal{C}_{6}=d_{0}\dprod \limits_{i<j=1}^{4}\left( s-t_{i}-t_{j}\right)
\end{equation}%
Notice that the breaking of monodromy induced by non trivial fluxes is
engineered by splitting spectral cover method as in eq(\ref{bb}) regarding
the spectral cover of $SU_{5}^{\bot }$. In the case of fundamental $\mathcal{%
C}_{4}$ of $SU_{4}^{\bot }$; we have quite similar decompositions; for
examples $\  \mathcal{C}_{4}=\mathcal{C}_{3}^{\left( 4\right) }\times 
\mathcal{C}_{1}^{\left( 4\right) }$ reducing $\mathbb{S}_{4}$ monodromy to $%
\mathbb{S}_{3};$ and $\mathcal{C}_{4}=\mathcal{C}_{2}^{\left( 4\right)
}\times \mathcal{C}_{2}^{\left( 4\right) }$ reducing $\mathbb{S}_{4}$
monodromy to $\mathbb{S}_{2}\times \mathbb{S}_{2}$.

\  \  \ 

$\bullet $ $SU_{5}\times \Gamma $ models\newline
An analogous description of $SO_{10}\times \Gamma $ models \ can done for
other F-theory GUTs. In the interesting case of the $SU_{5}\times \Gamma $
models the gauge symmetry is given by Georgi-Glashow group $SU_{5}$, and
monodromy groups $\Gamma $ contained in $\mathbb{S}_{5}$. A particular
branch of subgroups $\Gamma $ is the one contained in $\mathbb{S}_{4}$
listed in the following table, see fig. \ref{5}%
\begin{equation}
\begin{tabular}{lllll}
$\Gamma $ &  & order &  & multiplicity \\ \hline
$\  \mathbb{S}_{4}$ &  & $24$ &  & $\  \  \  \ 1$ \\ 
$\  \mathbb{A}_{4}$ &  & $12$ &  & $\  \  \  \ 1$ \\ 
$\  \mathbb{D}_{4}$ &  & $\ 8$ &  & $\  \  \  \ 3$ \\ 
$\  \mathbb{S}_{3}$ &  & $\ 6$ &  & $\  \  \  \ 4$ \\ 
$\  \mathbb{V}_{4}$ &  & $\ 4$ &  & $\  \  \  \ 1$ \\ 
$\  \mathbb{Z}_{4}$ &  & $\ 4$ &  & $\  \  \  \ 3$ \\ 
$\  \mathbb{Z}_{2}\times \mathbb{Z}_{2}$ &  & $\ 4$ &  & $\  \  \  \ 3$ \\ 
$\  \mathbb{Z}_{3}$ &  & $\ 3$ &  & $\  \  \  \ 4$ \\ 
$\  \mathbb{Z}_{2}$ &  & $\ 2$ &  & $\  \  \  \ 9$ \\ 
$\ I_{id}$ &  & $\ 1$ &  & $\  \  \  \ 1$ \\ \hline
\end{tabular}
\label{ls}
\end{equation}%
\begin{equation*}
\end{equation*}%
The elliptically fibered Calabi-Yau fourfolds $CY4\sim E_{{\tiny SU}_{{\tiny %
5}}}\times \mathcal{B}_{3}$ is locally realised by the Tate model; it is
described by the algebraic equation $y^{2}=x^{3}+b_{5}^{\prime
}xy+b_{4}^{\prime }x^{2}w+b_{3}^{\prime }yw^{2}+b_{2}^{\prime
}xw^{3}+b_{0}^{\prime }w^{5}$ with sections $b_{n}^{\prime }$ given by
holomorphic functions on $\mathcal{B}_{3}$ with properties as in eqs(\ref{w5}%
)-(\ref{5w}).

\  \  \  \  \  \newline
The matter curves of these models are read from the decomposition of the $%
\mathbf{248}$ adjoint representation of $E_{8}$ in terms of $SU_{5}\times
SU_{5}^{\bot }$ representations as given below%
\begin{equation}
\begin{tabular}{lll}
$\mathbf{248}$ & $\rightarrow $ & $\left( \mathbf{24},\mathbf{1}_{\perp
}\right) \oplus \left( \mathbf{1},\mathbf{24}_{\perp }\right) \oplus $ \\ 
&  & $\left( \mathbf{10},\mathbf{5}_{\perp }\right) \oplus \left( \mathbf{%
\bar{5}},\mathbf{10}_{\perp }\right) \oplus \left( \overline{\mathbf{10}},%
\mathbf{\bar{5}}_{\perp }\right) \oplus \left( \mathbf{5},\overline{\mathbf{%
10}}_{\perp }\right) $%
\end{tabular}%
\end{equation}%
In this $SU_{5}$ theory, the monodromy symmetry $\Gamma $ is contained in
the Weyl group of the perpendicular $SU_{5}^{\bot }$; and the matter content
of the model is labeled by five weights t$_{i}$ like%
\begin{equation}
\begin{tabular}{lllll}
$\mathbf{10}_{t_{i}},$ & $\overline{\mathbf{10}}_{-t_{i}},$ & $\mathbf{\bar{5%
}}_{t_{i}+t_{j}},$ & $\mathbf{5}_{-t_{i}-t_{j}},$ & $\mathbf{1}%
_{t_{i}-t_{j}} $%
\end{tabular}
\label{d}
\end{equation}%
with traceless condition%
\begin{equation}
t_{1}+t_{2}+t_{3}+t_{4}+t_{5}=0  \label{cd}
\end{equation}%
The spectral cover description of the matter curves (\ref{d}) is quite
similar to the above $SO_{10}\times \Gamma $ models one; for the fundamental
spectral cover $\mathcal{C}_{5}$ of $SU_{5}\times \mathbb{S}_{5}$ models,
see eqs(\ref{c5}-\ref{5c}). \newline
With these tools at hand, we are now in position to construct the fusion
algebras $\mathcal{F}_{\Gamma }$ of operators of the F- theory GUT spectrum $%
\left \{ \Phi _{R_{i}}\right \} $ by mainly focussing on $SO_{10}\times
\Gamma $ models by starting with largest $\Gamma =\mathbb{S}_{4}$; quite
similar constructions are valid for $SU_{5}\times \Gamma $ theory with $%
\Gamma \subset \mathbb{S}_{5}$.

\section{Fusion operators algebra with $\mathbb{S}_{4}$ symmetry}

To begin, notice that we have chosen to start by studying the fusion algebra 
$\mathcal{F}_{\mathbb{S}_{4}}$ of the set $\left \{ \Phi _{R_{i}}\right \} $
carrying quantum numbers in $\mathbb{S}_{4}$. Though the corresponding $%
SO_{10}\times \mathbb{S}_{4}$ model is not phenomenologically interesting
since only three matter generations are known and so $\mathbb{S}_{4}$ should
be broken down; we take the opportunity to illustrate how the $\Phi _{R_{i}}$%
's are involved in building superpotentials; and to derive the corresponding
fusion algebra $\mathcal{F}_{\mathbb{S}_{4}}$. \newline
To that purpose, we first give useful tools on $\mathbb{S}_{4}$
representations as involved in F-GUT; and turn after to build the $\mathbb{S}%
_{4}$- fusion algebra $\mathcal{F}_{\mathbb{S}_{4}}$.

\subsection{$\mathbb{S}_{4}$ as F-GUT monodromy}

The permutation symmetry $\mathbb{S}_{4}$ is a discrete group having \emph{24%
} elements arranged into \emph{5} conjugacy classes $\mathfrak{C}_{1},...,%
\mathfrak{C}_{5}$ as shown on table (\ref{1}) reported in appendix. It has 
\emph{5} irreducible representations $\boldsymbol{R}_{1},...,\boldsymbol{R}%
_{5}$ with dimensions $d_{i}$ given by the character relation linking the
order of $\mathbb{S}_{4}$ to the $\boldsymbol{R}_{i}$ dimensions like $%
24=\sum_{i}d_{i}^{2}$; by expanding we have 
\begin{equation}
24=1^{2}+\left( 1^{\prime }\right) ^{2}+2^{2}+3^{2}+\left( 3^{\prime
}\right) ^{2}
\end{equation}%
The $\mathbb{S}_{4}$ group has \emph{3} non commuting generators $a,$ $b,$ $%
c $ which can be chosen as respectively given by the 2- cycle $\left(
12\right) $, the 3-cycle $\left( 123\right) $ and the 4- cycle $\left(
1234\right) $. These generators obey amongst others the cyclic relations $%
a^{2}=b^{3}=c^{4}=I_{id}$; their characters $\mathbf{\chi }_{\boldsymbol{R}%
_{i}}{\small (a)}$, $\mathbf{\chi }_{\boldsymbol{R}_{i}}{\small (b)}$, $%
\mathbf{\chi }_{\boldsymbol{R}_{i}}{\small (c)}$ can be read from eq(\ref{1}%
); they are given by%
\begin{equation}
\begin{tabular}{|l|l|l|l|l|l|}
\hline
${\small \chi }_{\boldsymbol{R}_{i}}^{\alpha }$ & ${\small \chi }_{_{%
\boldsymbol{I}}}$ & ${\small \chi }_{_{\boldsymbol{3}}}$ & ${\small \chi }%
_{_{\boldsymbol{2}}}$ & ${\small \chi }_{_{\boldsymbol{3}^{\prime }}}$ & $%
{\small \chi }_{_{\epsilon }}$ \\ \hline
${\small a}$ & ${\small 1}$ & ${\small \  \ 1}$ & ${\small \  \ 0}$ & ${\small %
-1}$ & ${\small -1}$ \\ \hline
${\small b}$ & ${\small 1}$ & ${\small \  \ 0}$ & ${\small -1}$ & ${\small \
\ 0}$ & ${\small \  \ 1}$ \\ \hline
${\small c}$ & ${\small 1}$ & ${\small -1}$ & ${\small \  \ 0}$ & ${\small \
\ 1}$ & ${\small -1}$ \\ \hline
\end{tabular}
\label{a1}
\end{equation}%
The ${\small \chi }_{\boldsymbol{R}_{i}}^{\alpha }$ can be organised into 
\emph{5} character vectors $\mathbf{\vec{\chi}}_{i}=\left( \mathrm{\chi }%
_{i}^{a},\mathrm{\chi }_{i}^{b},\mathrm{\chi }_{i}^{c}\right) $ where we
have set $\mathrm{\chi }_{i}^{\alpha }=\mathrm{\chi }_{\boldsymbol{R}_{i}}%
{\small (\alpha )}$. Observe also the following remarkable property that
turns out to play an important role in the derivation of the fusion algebra
for $\mathbb{S}_{4}$ monodromy,%
\begin{equation}
\mathbf{\vec{\chi}}_{_{\boldsymbol{3}}}+\mathbf{\vec{\chi}}_{_{\boldsymbol{3}%
^{\prime }}}=\left( 0,0,0\right)  \label{1a}
\end{equation}%
These $\mathbf{\vec{\chi}}_{i}$'s will be used in this paper as a tool to
characterise the curves spectrum of $SO_{10}\times \mathbb{S}_{4}$ model.
Indeed, by following \textrm{\cite{1B,2B}}$\mathrm{,s}$ee also eqs(\ref{b}-%
\ref{c})\textrm{,} the matter curves in the spectrum of this supersymmetric
model involve three kinds of multiplets\textrm{\footnote{%
These multiplets are\ interpreted in the underlying low energy effective 4d $%
\mathcal{N}=1$ QFT in terms of chiral superfields $\Phi _{R_{i}}$ carrying
gauge quantum numbers; but also monodromy representations.}} transforming in
representations of $\mathbb{S}_{4}$; these are:

\begin{itemize}
\item the \emph{4} matter curves generally denoted as $\mathbf{16}_{\mu }$;
they describe the 16-plets of the $SO_{10}\times \mathbb{S}_{4}$ model;
three of them interpreted in terms of the usual GUT generations;

\item the \emph{6} Higgs curves $\mathbf{10}_{\left[ \mu \nu \right] }$
describing the 10-plets; the corresponding low energy (super) fields can
have VEVs giving masses to particles of GUT; and

\item the \emph{3}$+$\emph{12} curves $\mathbf{1}_{ij}$ describing the
flavons, denoted by $4\times 4$ traceless matrix $\vartheta _{\mu }^{\nu }$;
three of them neutral, and the \emph{12}$=6+6^{\prime }$, denoted below as $%
\tilde{\vartheta}_{\mu }^{\nu }$, are charged under $\mathbb{S}_{4}$; they
can have VEVs; they are important for generating mass hierarchy.
\end{itemize}

\  \  \  \newline
The curves spectrum of $SO_{10}\times \mathbb{S}_{4}$ model is commonly
presented as on following table%
\begin{equation}
\begin{tabular}{|c|c|c|c|}
\hline
{\small matters curves} & {\small weights} & {\small homology} & {\small U}$%
{\small (1)}_{{\small X}}${\small \ flux} \\ \hline
${\small 16}_{\mu }$ & ${\small t}_{\mu }$ & ${\small \eta -4c}_{1}$ & $%
{\small 0}$ \\ \hline
${\small 10}_{\left[ \mu \nu \right] }$ & $\left. \left( {\small t}_{\mu }%
{\small +t}_{\nu }\right) \right \vert _{{\small \mu <\nu }}$ & ${\small %
\eta }^{\prime }{\small -6c}_{1}$ & ${\small 0}$ \\ \hline
${\small \tilde{\vartheta}}_{\mu }^{\nu }$ & ${\small t}_{\mu }{\small -t}%
_{\nu }$ & ${\small \eta }^{\prime \prime }{\small -12c}_{1}$ & ${\small 0}$
\\ \hline
\end{tabular}
\label{a2}
\end{equation}%
where we have also given the homology classes (\ref{31}); whose
interpretations can be found in \textrm{\cite{4A,1B,2B}}; the last column is
trivial here; but it is important for the study of $SO_{10}\times \Gamma $
models with monodromy $\Gamma $ given by subgroups of $\mathbb{S}_{4}$.
There, the U$\left( 1\right) _{X}$ flux takes non zero values and splits the
spectral covers.

\subsection{Superpotentials and fusion algebra $\mathcal{F}_{\mathbb{S}_{4}}$%
}

To study the building of superpotentials $W\left( \Phi _{R_{i}}\right) $ of
low energy effective 4d $\mathcal{N}=1$ supersymmetric theory of $%
SO_{10}\times \mathbb{S}_{4}$ model, it is interesting to re-construct the
structure of table (\ref{a2}). By using the irreducible $\boldsymbol{R}_{i}$
representations of $\mathbb{S}_{4}$ as well as their $\mathbf{\chi }%
_{i}^{\alpha }$ characters (\ref{a1}), we can reformulate the curve spectrum
of $SO_{10}\times \mathbb{S}_{4}$- model as follows 
\begin{equation}
\begin{tabular}{|c|c|c|c|c|}
\hline
{\small matters curves} & $S_{4}${\small \ irreps} & {\small character }$%
\vec{\chi}_{i}$ & {\small homology} & {\small U}${\small (1)}_{{\small X}}$%
{\small \ flux} \\ \hline
$\left. 
\begin{array}{c}
{\small 16}_{{\small 0}} \\ 
{\small 16}_{i}%
\end{array}%
\right. $ & $\left. 
\begin{array}{c}
{\small 1}^{+} \\ 
{\small 3}^{+}%
\end{array}%
\right. $ & $\left. 
\begin{array}{c}
\left( {\small 1,1,1}\right) \\ 
\left( {\small 1,0,-1}\right)%
\end{array}%
\right. $ & $\left. 
\begin{array}{c}
{\small -c}_{1} \\ 
{\small \eta -3c}_{1}%
\end{array}%
\right. $ & $\left. 
\begin{array}{c}
{\small 0} \\ 
{\small 0}%
\end{array}%
\right. $ \\ \hline
$\left. 
\begin{array}{c}
{\small 10}_{i} \\ 
{\small 10}_{\left[ ij\right] }%
\end{array}%
\right. $ & $\left. 
\begin{array}{c}
{\small 3}^{+} \\ 
{\small 3}^{-}%
\end{array}%
\right. $ & $\left. 
\begin{array}{c}
\left( {\small 1,0,-1}\right) \\ 
\left( {\small -1,0,1}\right)%
\end{array}%
\right. $ & $\left. 
\begin{array}{c}
{\small \eta }^{\prime }{\small -3c}_{1} \\ 
{\small -3c}_{1}%
\end{array}%
\right. $ & $\left. 
\begin{array}{c}
{\small 0} \\ 
{\small 0}%
\end{array}%
\right. $ \\ \hline
$\left. 
\begin{array}{c}
{\small 1}_{i} \\ 
{\small 1}_{\left[ ij\right] }%
\end{array}%
\right. $ & $\left. 
\begin{array}{c}
{\small 3}^{+} \\ 
{\small 3}^{-}%
\end{array}%
\right. $ & $\left. 
\begin{array}{c}
\left( {\small 1,0,-1}\right) \\ 
\left( {\small -1,0,1}\right)%
\end{array}%
\right. $ & $\left. 
\begin{array}{c}
{\small \eta }^{\prime \prime }{\small -3c}_{1} \\ 
{\small -3c}_{1}%
\end{array}%
\right. $ & $\left. 
\begin{array}{c}
{\small 0} \\ 
{\small 0}%
\end{array}%
\right. $ \\ \hline
\end{tabular}
\label{a3}
\end{equation}%
where we have also used the reduction $\mathbf{12}=2\times \left( \mathbf{3}%
^{+}\oplus \mathbf{3}^{-}\right) $; see below for its derivation. We will
also use the convenient notations suggested by $\mathbb{S}_{4}$- characters (%
\ref{a2}),%
\begin{equation}
\begin{tabular}{llllllll}
$\mathbf{1}$ & $\equiv \mathbf{1}_{{\small (1,1,1)}}^{+}$ & , & $\mathbf{3}$
& $\equiv \mathbf{3}_{{\small (1,0,-1)}}^{+}$ & , & $\mathbf{2}$ & $\equiv 
\mathbf{2}_{{\small (0,-1,0)}}^{0}$ \\ 
$\mathbf{\epsilon }$ & $\equiv \mathbf{1}_{{\small (-1,1,-1)}}^{-}$ & , & $%
\mathbf{3}^{\prime }$ & $\equiv \mathbf{3}_{{\small (-1,0,1)}}^{-}$ & , &  & 
\end{tabular}
\label{b1}
\end{equation}

$\bullet $ \emph{Superpotentials }$W\left( \Phi _{R_{i}}\right) $\newline
Superpotentials $W\left( \Phi _{R_{i}}\right) $ of the low energy effective
QFT of $SO_{10}\times \mathbb{S}_{4}$- models are given by product of matter
chiral superfield operators $\Psi _{\mathbf{16}_{\mu }^{M}}\sim \mathbf{16}%
_{\mu }^{M}$, Higgs $\Psi _{\mathbf{10}_{\left[ \mu \nu \right] }^{H}}\sim 
\mathbf{10}_{\left[ \mu \nu \right] }^{H}$ and flavons $\Psi _{\mathbf{1}%
_{\alpha \beta }^{F}}\sim \mathbf{1}_{\alpha \beta }^{F}\equiv \mathbf{%
\vartheta }_{\alpha \beta }$. These superpotentials should be $SO_{10}$
gauge invariant; but also invariant under monodromy $\mathbb{S}_{4}$. A
typical example of a gauge invariant superpotential in $SO_{10}\times 
\mathbb{S}_{4}$- model is given by the tree level (top-quark) Yukawa
couplings 
\begin{equation}
W_{SO_{10}}^{tree}=\lambda _{{}}^{\mu \nu \rho \sigma }\mathbf{16}_{\mu
}^{M}\otimes \mathbf{16}_{\nu }^{M}\otimes \mathbf{10}_{\left[ \mu \nu %
\right] }^{H}  \label{27}
\end{equation}%
where $\lambda _{{}}^{\mu \nu \rho \sigma }$ are coupling constants
transforming as a rank 4 tensor; invariance under $\mathbb{S}_{4}$ puts
constraint on these $\lambda _{{}}^{\mu \nu \rho \sigma }$'s. Other forms of
superpotentials can be also written down; they involve flavons like in the
following non renormalisable 4-order one 
\begin{equation}
W_{SO_{10}}^{\left( 4\right) }=\lambda _{{}}^{\mu \nu \rho \sigma \alpha
\beta }\text{ }\mathbf{16}_{\mu }^{M}\otimes \mathbf{16}_{\nu }^{M}\otimes 
\mathbf{10}_{\left[ \mu \nu \right] }^{H}\otimes \mathbf{\vartheta }_{\left[
\alpha \beta \right] }^{F}  \label{w4}
\end{equation}%
In what follows, we develop a method to construct $\mathbb{S}_{4}$-
monodromy invariant superpotentials; this approach is based on the fusion
algebra $\mathcal{F}_{\mathbb{S}_{4}}$; itself based on the characters of
irreducible representations of $\mathbb{S}_{4}$. To derive $\mathcal{F}_{%
\mathbb{S}_{4}}$, we proceed as follows:

\begin{description}
\item[$\left( i\right) $] we start from the finite spectrum (\ref{a3}); and
denote the corresponding superfield operators by their representations and
characters like%
\begin{equation}
\begin{tabular}{|c|c|c|c|}
\hline
{\small matters curves} & ${\small SO}_{{\small 10}}$ & $S_{4}${\small \
irreps} & ${\small SO}_{10}{\small \times S}_{4}$ \\ \hline
$\left. 
\begin{array}{c}
{\small 16}_{{\small (1,1,1)}} \\ 
{\small 16}_{{\small (1,0,-1)}}%
\end{array}%
\right. $ & {\small 16} & $\left. 
\begin{array}{c}
{\small 1}^{+} \\ 
{\small 3}^{+}%
\end{array}%
\right. $ & $\left. 
\begin{array}{c}
{\small 16\otimes 1}_{{\small (1,1,1)}}^{+} \\ 
{\small 16\otimes 3}_{{\small (1,0,-1)}}^{+}%
\end{array}%
\right. $ \\ \hline
$\left. 
\begin{array}{c}
{\small 10}_{{\small (1,0,-1)}} \\ 
{\small 10}_{{\small (-1,0,1)}}%
\end{array}%
\right. $ & {\small 10} & $\left. 
\begin{array}{c}
{\small 3}^{+} \\ 
{\small 3}^{-}%
\end{array}%
\right. $ & $\left. 
\begin{array}{c}
{\small 10\otimes 3}_{{\small (1,0,-1)}}^{+} \\ 
{\small 10\otimes 3}_{{\small (-1,0,1)}}^{-}%
\end{array}%
\right. $ \\ \hline
$\left. 
\begin{array}{c}
{\small \vartheta }_{{\small (1,0,-1)}} \\ 
{\small \vartheta }_{{\small (-1,0,1)}}%
\end{array}%
\right. $ & {\small 1} & $\left. 
\begin{array}{c}
{\small 3}^{+} \\ 
{\small 3}^{-}%
\end{array}%
\right. $ & $\left. 
\begin{array}{c}
1{\small \otimes 3}_{{\small (1,0,-1)}}^{+} \\ 
{\small 1\otimes 3}_{{\small (-1,0,1)}}^{-}%
\end{array}%
\right. $ \\ \hline
\end{tabular}%
\end{equation}%
this means that the operator $\mathbf{16}_{{\small (1,1,1)}}$ is a trivial
singlet of $\mathbb{S}_{4};$ the $\mathbf{16}_{{\small (1,0,-1)}}$ is a $%
\mathbf{3}^{+}$- triplet of $\mathbb{S}_{4}$; the $\mathbf{10}_{{\small %
(-1,0,1)}}$ is a $\mathbf{3}^{-}$- triplet of $\mathbb{S}_{4}$; and so on.

\item[$\left( ii\right) $] seen that $\mathbf{16,}$ $\mathbf{10}$ and $%
\mathbf{\vartheta }$ are representations of $SO_{10}$; and seen that we are
interested in invariance under discrete $\mathbb{S}_{4}$, we will refer to
these superfield operators like%
\begin{equation}
\begin{tabular}{lllllllll}
$\mathbf{16}_{{\small (p}_{1}{\small ,q}_{1}{\small ,r}_{1}{\small )}}$ & ,
& $\mathbf{10}_{{\small (p}_{2}{\small ,q}_{2}{\small ,r}_{2}{\small )}}$ & ,
& $\mathbf{\vartheta }_{{\small (p}_{3}{\small ,q}_{3}{\small ,r}_{3}{\small %
)}}$ &  & $\rightarrow $ &  & $\mathbf{\Phi }_{{\small (p}_{i}{\small ,q}_{i}%
{\small ,r}_{i}{\small )}}$%
\end{tabular}%
\end{equation}%
In other words, think of $\mathbf{16}_{{\small (p}_{1}{\small ,q}_{1}{\small %
,r}_{1}{\small )}}$ as given by $\mathbf{16\otimes \Phi }_{{\small (p}_{1}%
{\small ,q}_{1}{\small ,r}_{1}{\small )}}$; and so on. Then focus on the
algebra of the $\mathbb{S}_{4}$ representations; the properties of the $SO_{%
{\small 10}}$ gauge multiplets are dealt as in \ usual GUT models.

\item[$\left( iii\right) $] Generic superpotentials $W\left( \Phi
_{R_{i}}\right) $ in $\mathbb{S}_{4}$- model are given by 
\begin{equation}
W_{SO_{10}}=\sum tr\left[ \mathcal{R}_{\left( {\small P,Q,R}\right) }\right]
\label{11}
\end{equation}%
with%
\begin{equation}
\mathcal{R}_{\left( {\small P,Q,R}\right) }=\Phi _{{\small (p}_{1}{\small ,q}%
_{1}{\small ,r}_{1}{\small )}}\otimes \cdots \otimes \Phi _{{\small (p}_{n}%
{\small ,q}_{n}{\small ,r}_{n}{\small )}}
\end{equation}%
where the trace refers both to invariance under gauge symmetry; and $\mathbb{%
S}_{4}$ monodromy. An example of $\mathcal{R}_{\left( P,Q,R\right) }$ is
given by the Yukawa tri-coupling 
\begin{equation}
W_{SO_{10}}^{tree}=\mathbf{16}_{{\small (p}_{1}{\small ,q}_{1}{\small ,r}_{1}%
{\small )}}^{M}\otimes \mathbf{16}_{{\small (p}_{2}{\small ,q}_{2}{\small ,r}%
_{2}{\small )}}^{M}\otimes \mathbf{10}_{{\small (p}_{3}{\small ,q}_{3}%
{\small ,r}_{3}{\small )}}^{H}  \label{w3}
\end{equation}%
If we take all these \emph{3} operators as $\mathbb{S}_{4}$- triplets like 
\begin{equation}
\begin{tabular}{lll}
$\mathbf{16}_{{\small (p}_{1}{\small ,q}_{1}{\small ,r}_{1}{\small )}}^{M}$
& $=$ & $\mathbf{16}_{{\small (1,0,-1)}}$ \\ 
$\mathbf{16}_{{\small (p}_{2}{\small ,q}_{2}{\small ,r}_{2}{\small )}}^{M}$
& $=$ & $\mathbf{16}_{{\small (1,0,-1)}}$ \\ 
$\mathbf{10}_{{\small (p}_{3}{\small ,q}_{3}{\small ,r}_{3}{\small )}}^{H}$
& $=$ & $\mathbf{10}_{{\small (-1,0,1)}}$%
\end{tabular}%
\end{equation}%
we end with the following reducible 27-$\dim $ representation of $\mathbb{S}%
_{4}$%
\begin{equation}
\mathbf{\Phi }_{{\small (1,0,-1)}}\otimes \mathbf{\Phi }_{{\small (1,0,-1)}%
}^{\prime }\otimes \mathbf{\Phi }_{{\small (-1,0,1)}}^{\prime \prime
}=\sum_{\left( {\small p,q,r}\right) }n_{\left( {\small p,q,r}\right) }%
\boldsymbol{R}_{\left( {\small p,q,r}\right) }
\end{equation}%
where the positive integers $n_{\left( {\small p,q,r}\right) }$ are some
multiplicities $n_{\left( {\small p,q,r}\right) }$ constrained by the total
dimension of the tensor product of representations. Other examples of chiral
superpotentials are given by higher order superpotentials involving flavons;
for instance $W\left( \Phi _{R_{i}}\right)
=W_{SO_{10}}^{tree}+W_{SO_{10}}^{\left( 4\right) }$ given by the sum $(\ref%
{27})+(\ref{w4})$.
\end{description}

\  \  \ 

$\bullet $ \emph{Fusion algebra }$\mathcal{F}_{\mathbb{S}_{4}}$\newline
To compute the explicit expression of (\ref{11}), one needs reducing the
tensor product $\mathcal{R}_{\left( P,Q,R\right) }$ in terms of a direct sum
over the \emph{5} irreducible representations of $\mathbb{S}_{4}$ as follows 
\begin{equation}
\mathcal{R}_{\left( P,Q,R\right) }=n_{e}\mathbf{1}_{{\small (1,1,1)}%
}^{+}\oplus n_{\epsilon }\mathbf{1}_{{\small (-1,1,-1)}}^{-}\oplus n_{2}%
\mathbf{2}_{{\small (0,-1,0)}}^{0}\oplus n_{+}\mathbf{3}_{{\small (1,0,-1)}%
}^{+}\oplus n_{-}\mathbf{3}_{{\small (-1,0,1)}}^{-}
\end{equation}%
The $n_{i}$'s are obtained by demanding two conservation laws; total
dimension and total character. But to fully achieve the reduction of (\ref%
{11}), one must know the fusion algebra of two operators $\Phi _{{\small (p}%
_{i}{\small ,q}_{i}{\small ,r}_{i}{\small )}}\otimes \Phi _{{\small (p}_{j}%
{\small ,q}_{j}{\small ,r}_{j}{\small )}}$; then proceed step by step until
getting the full reduction as above. In other words, it is enough to know
the right hand of the following expansion%
\begin{equation}
\Phi _{{\small (p}_{i}{\small ,q}_{i}{\small ,r}_{i}{\small )}}\otimes \Phi
_{{\small (p}_{j}{\small ,q}_{j}{\small ,r}_{j}{\small )}}=\sum_{k}C_{\left%
\{ {\small (p}_{i}{\small ,q}_{i}{\small ,r}_{i}\right \} ,\left \{ {\small p%
}_{j}{\small ,q}_{j}{\small ,r}_{j}\right \} }^{\left \{ {\small p}_{k}%
{\small ,q}_{k}{\small ,r}_{k}\right \} }\Phi _{{\small (p}_{k}{\small ,q}%
_{k}{\small ,r}_{k}{\small )}}
\end{equation}%
By using the irreducible $\boldsymbol{R}_{i}$ representations of $\mathbb{S}%
_{4}$, the above fusion equation reads in a condensed manner as follows 
\begin{equation}
\Phi _{\boldsymbol{R}_{i}}\otimes \Phi _{\boldsymbol{R}_{j}}=\sum_{%
\boldsymbol{R}_{k}}C_{\boldsymbol{R}_{i},\boldsymbol{R}_{j}}^{\boldsymbol{R}%
_{k}}\Phi _{\boldsymbol{R}_{k}}
\end{equation}%
or formally like%
\begin{equation}
\boldsymbol{R}_{i}\otimes \boldsymbol{R}_{j}=\oplus _{\boldsymbol{R}_{k}}C_{%
\boldsymbol{R}_{i},\boldsymbol{R}_{j}}^{\boldsymbol{R}_{k}}\boldsymbol{R}_{k}
\end{equation}%
As one of the results of this paper is that the fusion algebras $\mathcal{F}%
_{\mathbb{S}_{4}}$ of the $\mathbb{S}_{4}$ monodromy symmetry is given, in
addition to $\mathbf{1}_{{\small (1,1,1)}}^{+}\otimes \boldsymbol{R}_{%
{\small (p,q,r)}}=\boldsymbol{R}_{{\small (p,q,r)}}$, by the following
relations preserving dimensions and characters%
\begin{equation}
\ 
\begin{tabular}{lllll}
$\mathbf{3}^{\pm }$ & $\otimes $ & $\mathbf{3}^{\pm }$ & $=$ & $\mathbf{1}%
^{+}\oplus \mathbf{2}^{0}\oplus \mathbf{3}^{+}\oplus \mathbf{3}^{-}$ \\ 
$\mathbf{3}^{\pm }$ & $\otimes $ & $\mathbf{2}^{0}$ & $=$ & $\mathbf{3}%
^{+}\oplus \mathbf{3}^{-}$ \\ 
$\mathbf{3}^{\pm }$ & $\otimes $ & $\mathbf{1}^{-}$ & $=$ & $\mathbf{3}^{\mp
}$ \\ 
$\mathbf{3}^{+}$ & $\otimes $ & $\mathbf{3}^{-}$ & $=$ & $\mathbf{1}%
^{-}\oplus \mathbf{2}^{0}\oplus \mathbf{3}^{+}\oplus \mathbf{3}^{-}$ \\ 
$\mathbf{2}^{0}$ & $\otimes $ & $\mathbf{2}^{0}$ & $=$ & $\mathbf{1}%
^{+}\oplus \mathbf{1}^{-}\oplus \mathbf{2}^{0}$ \\ 
$\mathbf{2}^{0}$ & $\otimes $ & $\mathbf{1}^{-}$ & $=$ & $\mathbf{2}^{0}$ \\ 
$\mathbf{1}^{-}$ & $\otimes $ & $\mathbf{1}^{-}$ & $=$ & $\mathbf{1}^{+}$%
\end{tabular}
\label{F}
\end{equation}%
By using (\ref{b1}), these relations read into a condensed manner like%
\begin{equation}
\begin{tabular}{lllll}
$\mathbf{3}_{{\small (1,0,-1)}}^{+}$ & $\  \otimes $ & $\  \mathbf{3}_{{\small %
(1,0,-1)}}^{+}$ & $\ =$ \  & $\  \mathbf{9}_{{\small (1,0,1)}}^{+}$ \\ 
$\mathbf{3}_{{\small (1,0,-1)}}^{+}$ & $\  \otimes $ & $\  \mathbf{3}_{{\small %
(-1,0,1)}}^{-}$ & $\ =$ \  & $\  \mathbf{9}_{{\small (-1,0,-1)}}^{-}$ \\ 
$\mathbf{3}_{{\small (1,0,-1)}}^{+}$ & $\  \otimes $ & $\  \  \mathbf{2}_{%
{\small (0,-1,0)}}^{0}$ & $\ =$ \  & $\  \mathbf{6}_{{\small (0,0,0)}}^{0}$
\\ 
$\mathbf{3}_{{\small (1,0,-1)}}^{+}$ & $\  \otimes $ & $\  \  \mathbf{1}_{%
{\small (-1,1,-1)}}^{-}$ & $\ =$ \  & $\  \mathbf{3}_{{\small (-1,0,1)}}^{-}$
\\ 
$\mathbf{3}_{{\small (-1,0,1)}}^{-}$ & $\  \otimes $ & $\  \mathbf{3}_{{\small %
(-1,0,1)}}^{-}$ & $\ =$ \  & $\  \mathbf{9}_{{\small (1,0,1)}}^{+}$ \\ 
$\mathbf{3}_{{\small (-1,0,1)}}^{-}$ & $\  \otimes $ & $\  \  \mathbf{2}_{%
{\small (0,-1,0)}}^{0}$ & $\ =$ \  & $\  \mathbf{6}_{{\small (0,0,0)}}^{0}$
\\ 
$\mathbf{2}_{{\small (0,-1,0)}}^{0}$ & $\  \otimes $ & $\  \  \mathbf{2}_{%
{\small (0,-1,0)}}^{0}$ & $\ =$ \  & $\  \mathbf{4}_{{\small (0,1,0)}}^{0}$
\\ 
$\mathbf{2}_{{\small (0,-1,0)}}^{0}$ & $\  \otimes $ & $\  \  \mathbf{1}_{%
{\small (-1,1,-1)}}^{-}$ & $\ =$ \  & $\  \mathbf{2}_{{\small (0,-1,0)}}^{0}$
\\ 
$\mathbf{1}_{{\small (-1,1,-1)}}^{-}$ & $\  \otimes $ & $\  \  \mathbf{1}_{%
{\small (-1,1,-1)}}^{-}$ & $\ =$ \  & $\  \mathbf{1}_{{\small (1,1,1)}}^{+}$%
\end{tabular}
\label{c1}
\end{equation}%
with right hand side given by the following%
\begin{equation}
\begin{tabular}{lllllll}
$\mathbf{9}_{{\small (1,0,1)}}^{+}$ & $=$ & $\mathbf{6}_{{\small (0,0,0)}%
}^{0}$ & $\oplus $ & $\mathbf{2}_{{\small (0,-1,0)}}^{0}$ & $\oplus $ & $%
\mathbf{1}_{{\small (1,1,1)}}^{+}$ \\ 
$\mathbf{9}_{{\small (-1,0,-1)}}^{-}$ & $=$ & $\mathbf{6}_{{\small (0,0,0)}%
}^{0}$ & $\oplus $ & $\mathbf{2}_{{\small (0,-1,0)}}^{0}$ & $\oplus $ & $%
\mathbf{1}_{{\small (-1,1,-1)}}^{-}$ \\ 
$\mathbf{6}_{{\small (0,0,0)}}^{0}$ & $=$ & $\mathbf{3}_{{\small (1,0,-1)}%
}^{+}$ & $\oplus $ & $\mathbf{3}_{{\small (-1,0,1)}}^{-}$ &  &  \\ 
$\mathbf{4}_{{\small (0,1,0)}}^{0}$ & $=$ & $\mathbf{2}_{{\small (0,-1,0)}%
}^{0}$ & $\oplus $ & $\mathbf{1}_{{\small (-1,1,-1)}}^{-}$ & $\oplus $ & $%
\mathbf{1}_{{\small (1,1,1)}}^{+}$%
\end{tabular}
\label{c2}
\end{equation}%
By substituting $\mathbf{6}_{{\small (0,0,0)}}^{0}=\mathbf{3}_{{\small %
(1,0,-1)}}^{+}\oplus \mathbf{3}_{{\small (-1,0,1)}}^{-}$ back into $\mathbf{9%
}_{{\small (1,0,1)}}^{\pm }$; we can read the $n_{\left( {\small p,q,r}%
\right) }$ multiplicities of the irreducible representations (\ref{b1}) of $%
\mathbb{S}_{4}$ monodromy 
\begin{equation}
\begin{tabular}{lllllllll}
$\mathbf{9}_{{\small (1,0,1)}}^{+}$ & $=$ & $\mathbf{3}_{{\small (1,0,-1)}%
}^{+}$ & $\oplus $ & $\mathbf{3}_{{\small (-1,0,1)}}^{-}$ & $\oplus $ & $%
\mathbf{2}_{{\small (0,-1,0)}}^{0}$ & $\oplus $ & $\mathbf{1}_{{\small %
(1,1,1)}}^{+}$ \\ 
$\mathbf{9}_{{\small (-1,0,-1)}}^{-}$ & $=$ & $\mathbf{3}_{{\small (1,0,-1)}%
}^{+}$ & $\oplus $ & $\mathbf{3}_{{\small (-1,0,1)}}^{-}$ & $\oplus $ & $%
\mathbf{2}_{{\small (0,-1,0)}}^{0}$ & $\oplus $ & $\mathbf{1}_{{\small %
(-1,1,-1)}}^{-}$%
\end{tabular}%
\end{equation}%
From the operators fusion algebra (\ref{F}), we learn that $\mathbf{9}_{%
{\small (1,0,1)}}^{+}$ and $\mathbf{4}_{{\small (0,1,0)}}^{0}$ have $\mathbb{%
S}_{4}$- monodromy invariants $\mathbf{1}_{{\small (1,1,1)}}^{+}$; while $%
\mathbf{9}_{{\small (-1,0,-1)}}^{-}$ and $\mathbf{6}_{{\small (0,0,0)}}^{0}$
haven't. The explicit derivation of $\mathcal{F}_{\mathbb{S}_{4}}$ is
straightforward; it relies on requiring both sides of (\ref{F}) to have same
representation character and same dimension; these properties have been
explicitly exhibited on (\ref{c1}-\ref{c2}).

\section{Fusion algebras $\mathcal{F}_{\mathbb{A}_{4}}$, $\mathcal{F}_{%
\mathbb{D}_{4}}$, $\mathcal{F}_{\mathbb{S}_{3}}$}

In this section, we extend the construction of (\ref{F}-\ref{c2}) to non
abelian subgroups of $\mathbb{S}_{4}$; first we consider the alternating
subgroup $\mathbb{A}_{4}$; then $\mathbb{S}_{3}$ and after the dihedral $%
\mathbb{D}_{4}$.

\subsection{Algebra with $\mathbb{A}_{4}$ monodromy}

The group $\mathbb{A}_{4}$ is the $\mathbb{S}_{4}$- subgroup of even
permutations; it has \emph{12} elements arranged into \emph{4} conjugacy
classes $\mathfrak{C}_{1},...,\mathfrak{C}_{4}$ as on table (\ref{22}) of
appendix; it has \emph{4} irreducible representations $\boldsymbol{R}%
_{1},...,\boldsymbol{R}_{4}$ with dimensions as in 
\begin{equation}
12=1^{2}+\left( 1^{\prime }\right) ^{2}+\left( 1^{\prime \prime }\right)
^{2}+3^{2}  \label{41}
\end{equation}%
Non abelian $\mathbb{A}_{4}$ has \emph{2} generators $\alpha $ and $\beta $
with characters like%
\begin{equation}
\begin{tabular}{|l|l|l|l|l|}
\hline
$\mathbb{A}_{4}$ & ${\small \  \  \chi }_{_{\boldsymbol{1}}}$ & ${\small \  \
\chi }_{_{\boldsymbol{1}^{\prime }}}$ & ${\small \  \  \chi }_{_{\boldsymbol{1}%
^{\prime \prime }}}$ & ${\small \  \  \chi }_{_{\boldsymbol{3}}}$ \\ \hline
${\small \alpha }$ & ${\small \  \ 1}$ & ${\small \  \ 1}$ & ${\small \  \ 1}$
& ${\small -1}$ \\ \hline
${\small \beta }$ & ${\small \  \ 1}$ & ${\small \  \ j}$ & ${\small \  \ j}%
^{2} $ & ${\small \  \ 0}$ \\ \hline
\end{tabular}%
\end{equation}%
with $j^{3}=1$. By denoting the irreducible representations of $\mathbb{A}%
_{4}$ as%
\begin{equation}
\begin{tabular}{lllll}
$\mathbf{1}$ & $\equiv \mathbf{1}_{{\small (1,1)}}^{0}$ & , & $\mathbf{3}$ & 
$\equiv \mathbf{3}_{{\small (-1,0)}}^{0}$ \\ 
$\mathbf{1}^{\prime }$ & $\equiv \mathbf{1}_{{\small (1,j)}}^{+}$ & , & $%
\mathbf{1}^{\prime \prime }$ & $\equiv \mathbf{1}_{{\small (1,j^{2})}}^{-}$%
\end{tabular}%
\end{equation}%
we find that the fusion algebra $\mathcal{F}_{\mathbb{A}_{4}}$ preserving
dimensions and characters is given by%
\begin{equation}
\ 
\begin{tabular}{lllll}
$\mathbf{3}_{{\small (-1,0)}}^{0}$ & $\otimes $ & $\mathbf{3}_{{\small (-1,0)%
}}^{0}$ & $=$ & $\mathbf{9}_{{\small (1,0)}}^{0}$ \\ 
$\mathbf{3}_{{\small (-1,0)}}^{0}$ & $\otimes $ & $\mathbf{1}_{\left(
1,j^{q}\right) }^{q}$ & $=$ & $\mathbf{3}_{{\small (-1,0)}}^{0}$ \\ 
$\mathbf{1}_{\left( 1,j^{q}\right) }^{q}$ & $\otimes $ & $\mathbf{1}_{\left(
1,j^{p}\right) }^{p}$ & $=$ & $\mathbf{1}_{\left( 1,j^{p+q}\right) }^{q+p}$%
\end{tabular}
\label{f4}
\end{equation}%
with%
\begin{equation}
\mathbf{9}_{{\small (1,0)}}^{0}=\mathbf{1}_{{\small (1,1)}}^{0}\oplus 
\mathbf{1}_{{\small (1,j)}}^{+}\oplus \mathbf{1}_{{\small (1,j^{2})}%
}^{-}\oplus \mathbf{3}_{{\small (-1,0)}}^{0}\oplus \mathbf{3}_{{\small (-1,0)%
}}^{0}
\end{equation}%
where the three singlets appear once; and the triplet twice.

\subsection{Fusion algebra $\mathcal{F}_{\mathbb{S}_{3}}$}

The order 6 group $\mathbb{S}_{3}$ has \emph{3} conjugacy classes $\mathfrak{%
C}_{1}$, $\mathfrak{C}_{2}$, $\mathfrak{C}_{3};$ and \emph{3} irreducible
representations $\boldsymbol{R}_{1},$ $\boldsymbol{R}_{2},$ $\boldsymbol{R}%
_{3}$ as reported in (\ref{33}) with dimensions read from the following
relation 
\begin{equation}
6=1^{2}+\left( 1^{\prime }\right) ^{2}+2^{2}  \label{42}
\end{equation}%
This finite group has two non commuting generators $\mathbf{a}$ and $\mathbf{%
b}$ with characters as follows%
\begin{equation}
\begin{tabular}{|l|l|l|l|}
\hline
${\small \chi }_{_{\boldsymbol{R}_{i}}}^{g}$ & ${\small \chi }_{_{%
\boldsymbol{I}}}$ & ${\small \chi }_{_{\boldsymbol{2}}}$ & ${\small \chi }%
_{_{\epsilon }}$ \\ \hline
${\small a}$ & ${\small \  \ 1}$ & ${\small \  \ 0}$ & ${\small -1}$ \\ \hline
${\small b}$ & ${\small \  \ 1}$ & ${\small -1}$ & ${\small \  \ 1}$ \\ \hline
\end{tabular}
\label{ch}
\end{equation}%
Denoting the three representations like%
\begin{equation}
\begin{tabular}{llllllll}
$\mathbf{1}$ & $\equiv \mathbf{1}_{{\small (1,1)}}^{+}$ & , & $\mathbf{1}%
^{\prime }$ & $\equiv \mathbf{1}_{{\small (-1,1)}}^{-}$ & , & $\mathbf{2}$ & 
$\equiv \mathbf{2}_{{\small (0,-1)}}^{0}$%
\end{tabular}%
\end{equation}%
we find that the fusion algebra $\mathcal{F}_{\mathbb{S}_{3}}$ preserving
dimension and character is given by%
\begin{equation}
\begin{tabular}{lllll}
$\mathbf{2}_{{\small (0,-1)}}^{0}$ & $\otimes $ & $\mathbf{2}_{{\small (0,-1)%
}}^{0}$ & $=$ & $\mathbf{4}_{{\small (0,1)}}^{0}$ \\ 
$\mathbf{2}_{{\small (0,-1)}}^{0}$ & $\otimes $ & $\mathbf{1}_{{\small (-1,1)%
}}^{-}$ & $=$ & $\mathbf{2}_{{\small (0,-1)}}^{-}$ \\ 
$\mathbf{1}_{{\small (-1,1)}}^{-}$ & $\otimes $ & $\mathbf{1}_{{\small (-1,1)%
}}^{-}$ & $=$ & $\mathbf{1}_{{\small (1,1)}}^{+}$%
\end{tabular}
\label{ss3}
\end{equation}%
with%
\begin{equation}
\mathbf{4}_{{\small (0,1)}}^{0}=\mathbf{1}_{{\small (1,1)}}^{+}\oplus 
\mathbf{1}_{{\small (-1,1)}}^{-}\oplus \mathbf{2}_{{\small (0,-1)}}^{0}
\end{equation}%
Notice that $\mathcal{F}_{\mathbb{S}_{3}}$ is a subalgebra of $\mathcal{F}_{%
\mathbb{S}_{4}}$; this can be seen by comparing (\ref{ss3}) with three last
rows of (\ref{F}).

\subsection{Dihedral $\mathbb{D}_{4}$ symmetry}

The $\mathbb{D}_{4}$ is an order \emph{8} subgroup of $\mathbb{S}_{4}$; it
has \emph{5} irreducible representations as on 
\begin{equation}
8=\left( 1_{1}\right) ^{2}+\left( 1_{2}\right) ^{2}+\left( 1_{3}\right)
^{2}+\left( 1_{4}\right) ^{2}+2^{2}  \label{43}
\end{equation}%
it has \emph{2} non commuting generators $a,c,$ satisfying $a^{2}=1$, $%
c^{4}=1,$ and $aca^{-1}=c^{-1}$; and \emph{5} conjugation classes%
\begin{equation}
\begin{tabular}{lllll}
$\mathfrak{C}_{1}\equiv \left \{ e\right \} $ & , & $\mathfrak{C}_{2}\equiv
\left \{ c^{2}\right \} $ & , & $\mathfrak{C}_{3}\equiv \left \{
c,c^{3}\right \} $ \\ 
$\mathfrak{C}_{4}\equiv \left \{ a,c^{2}a\right \} $ & $,$ & $\mathfrak{C}%
_{5}\equiv \left \{ ca,c^{3}a\right \} $ &  & 
\end{tabular}%
\end{equation}%
with character table as; see also appendix eq(\ref{hc}), 
\begin{equation}
\begin{tabular}{|l|l|l|l|l|l|}
\hline
${\small \chi }_{_{\boldsymbol{R}_{i}}}^{g}$ & $\  \  \mathrm{\chi }_{_{%
\mathbf{1}_{1}}}$ & $\  \  \mathrm{\chi }_{_{\mathbf{1}_{2}}}$ & $\  \  \mathrm{%
\chi }_{_{\mathbf{1}_{3}}}$ & $\  \  \mathrm{\chi }_{_{\mathbf{1}_{4}}}$ & $\
\  \mathrm{\chi }_{_{2}}$ \\ \hline
$a$ & $\  \ 1$ & $-1$ & $\  \ 1$ & $-1$ & $\  \ 0$ \\ \hline
$c$ & $\  \ 1$ & $\  \ 1$ & $-1$ & $-1$ & $\  \ 0$ \\ \hline
\end{tabular}%
\end{equation}%
Denoting the \emph{5} irreducible representations like%
\begin{equation}
\begin{tabular}{llllllll}
$\mathbf{1}_{1}$ & $\equiv \mathbf{1}_{{\small (1,1)}}$ & , & $\mathbf{1}%
_{3} $ & $\equiv \mathbf{1}_{{\small (1,-1)}}$ & , & $\mathbf{2}$ & $\equiv 
\mathbf{2}_{{\small (0,0)}}$ \\ 
$\mathbf{1}_{2}$ & $\equiv \mathbf{1}_{{\small (-1,1)}}$ & , & $\mathbf{1}%
_{4}$ & $\equiv \mathbf{1}_{{\small (-1,-1)}}$ & , &  & 
\end{tabular}%
\end{equation}%
and solving the conditions for $\mathbb{D}_{4}$- fusion algebra $\mathcal{F}%
_{\mathbb{D}_{4}}$ preserving dimension and character; we find 
\begin{equation}
\begin{tabular}{lllll}
$\mathbf{2}_{{\small (0,0)}}$ & $\otimes $ & $\mathbf{2}_{{\small (0,0)}}$ & 
$=$ & $\mathbf{4}_{{\small (0,0)}}$ \\ 
$\mathbf{2}_{{\small (0,0)}}$ & $\otimes $ & $\mathbf{1}_{{\small (p,q)}}$ & 
$=$ & $\mathbf{2}_{{\small (0,0)}}$ \\ 
$\mathbf{1}_{{\small (p,q)}}$ & $\otimes $ & $\mathbf{1}_{{\small (p}%
^{\prime }{\small ,q}^{\prime }{\small )}}$ & $=$ & $\mathbf{1}_{{\small (pp}%
^{\prime }{\small ,qq}^{\prime }{\small )}}$%
\end{tabular}%
\end{equation}%
with the four following solutions%
\begin{equation}
\begin{tabular}{llllllllllll}
$\mathcal{F}_{\mathbb{D}_{4}}^{{\small (I)}}$ & $:$ &  & $\mathbf{4}_{%
{\small (0,0)}}$ & $=$ & $\mathbf{1}_{{\small (1,1)}}$ & $\oplus $ & $%
\mathbf{1}_{{\small (-1,-1)}}$ & $\oplus $ & $\mathbf{1}_{{\small (1,-1)}}$
& $\oplus $ & $\mathbf{1}_{{\small (-1,1)}}$ \  \\ 
&  &  &  &  &  &  &  &  &  &  &  \\ 
$\mathcal{F}_{\mathbb{D}_{4}}^{{\small (II)}}$ & $:$ &  & $\mathbf{4}_{%
{\small (0,0)}}$ & $=$ & $\mathbf{1}_{{\small (1,1)}}$ & $\oplus $ & $%
\mathbf{1}_{{\small (-1,-1)}}$ & $\oplus $ & $\mathbf{2}_{{\small (0,0)}}$ & 
&  \\ 
&  &  &  &  &  &  &  &  &  &  &  \\ 
$\mathcal{F}_{\mathbb{D}_{4}}^{{\small (III)}}$ & $:$ &  & $\mathbf{4}_{%
{\small (0,0)}}$ & $=$ & $\mathbf{1}_{{\small (1,-1)}}$ & $\oplus $ & $%
\mathbf{1}_{{\small (-1,1)}}$ & $\oplus $ & $\mathbf{2}_{{\small (0,0)}}$ & 
&  \\ 
&  &  &  &  &  &  &  &  &  &  &  \\ 
$\mathcal{F}_{\mathbb{D}_{4}}^{{\small (IV)}}$ & $:$ &  & $\mathbf{4}_{%
{\small (0,0)}}$ & $=$ & $\mathbf{2}_{{\small (0,0)}}$ & $\oplus $ & $%
\mathbf{2}_{{\small (0,0)}}$ &  &  &  & 
\end{tabular}%
\end{equation}%
These relations teach us that generally speaking there are four fusion
algebras $\mathcal{F}_{\mathbb{D}_{4}}$.

\section{Extension to higher monodromies}

In SU$_{5}\times \Gamma $ models; monodromies are contained in $\mathbb{S}%
_{5}$; here we give two extensions of $\mathcal{F}_{\mathbb{S}_{4}}$; we
first give $\mathcal{F}_{\mathbb{S}_{5}}$, and then $\mathcal{F}_{\mathbb{A}%
_{5}}$.

\subsection{Fusion algebra $\mathcal{F}_{\mathbb{S}_{5}}$}

The group $\mathbb{S}_{5}$ has \emph{120} elements arranged into \emph{7}
conjugacy classes $\mathfrak{C}_{i}$ as on (\ref{2}); \emph{7} irreducible
representations $\boldsymbol{R}_{i}$ with dimensions as in the expansion 
\begin{equation}
120=1^{2}+\left( 1^{\prime }\right) ^{2}+4^{2}+\left( 4^{\prime }\right)
^{2}+5^{2}+\left( 5^{\prime }\right) ^{2}+6^{2}
\end{equation}%
It has \emph{4} non commuting generators $a,$ $b,$ $c,$ $d$ which can be
chosen as $\left( 12\right) $, $\left( 123\right) ,$ $\left( 1234\right) ,$ $%
\left( 12345\right) $; they obey amongst others the cyclic $%
a^{2}=b^{3}=c^{4}=d^{2}=I_{id}$; their characters are as follows%
\begin{equation}
\begin{tabular}{|l|l|l|l|l|l|l|l|}
\hline
${\small \chi }_{_{\boldsymbol{R}_{i}}}^{g}$ & ${\small \  \  \chi }_{_{1}}$ & 
${\small \  \  \chi }_{_{1^{\prime }}}$ & ${\small \  \  \chi }_{_{4}}$ & $%
{\small \  \  \chi }_{_{4^{\prime }}}$ & ${\small \  \  \chi }_{_{5}}$ & $%
{\small \  \  \chi }_{_{5^{\prime }}}$ & ${\small \  \  \chi }_{_{6}}$ \\ \hline
${\small a}$ & ${\small \  \ 1}$ & ${\small -1}$ & ${\small \  \ 2}$ & $%
{\small -2}$ & ${\small \  \ 1}$ & ${\small -1}$ & ${\small \  \ 0}$ \\ \hline
${\small b}$ & ${\small \  \ 1}$ & ${\small \  \ 1}$ & ${\small \  \ 1}$ & $%
{\small \  \ 1}$ & ${\small -1}$ & ${\small -1}$ & ${\small \ 0}$ \\ \hline
${\small c}$ & ${\small \  \ 1}$ & ${\small -1}$ & ${\small \  \ 0}$ & $%
{\small \  \ 0}$ & ${\small -1}$ & ${\small \  \ 1}$ & ${\small \ 0}$ \\ \hline
${\small d}$ & ${\small \  \ 1}$ & ${\small \  \ 1}$ & ${\small -1}$ & $%
{\small -1}$ & ${\small \ 0}$ & ${\small \  \ 0}$ & ${\small \ 1}$ \\ \hline
\end{tabular}%
\end{equation}%
We denote the \emph{7} irreducible representations like 
\begin{equation}
\begin{tabular}{lllllllll}
$\mathbf{1}=\boldsymbol{1}_{{\small (1,1,1,1)}}^{+}$ &  & , &  & $\mathbf{1}%
^{\prime }=\boldsymbol{1}_{{\small (-1,1,-1,1)}}^{-}$ &  & , &  & $\mathbf{6}%
=\boldsymbol{6}_{{\small (0,0,0,1)}}^{0}$ \\ 
$\mathbf{4}=\boldsymbol{4}_{{\small (2,1,0,-1)}}^{+}$ &  & , &  & $\mathbf{4}%
^{\prime }=\boldsymbol{4}_{{\small (-2,1,0,-1)}}^{-}$ &  &  &  &  \\ 
$\mathbf{5}=\boldsymbol{5}_{{\small (1,-1,-1,0)}}^{+}$ &  & , &  & $\mathbf{5%
}^{\prime }=\boldsymbol{5}_{{\small (-1,-1,1,0)}}^{-}$ &  &  &  & 
\end{tabular}%
\end{equation}%
The fusion algebra $\mathcal{F}_{\mathbb{S}_{5}}$ preserving dimensions and
characters is big but closed; it reads in a condensed manner as follows:%
\begin{equation}
\begin{tabular}{lllll}
$\mathbf{6}_{{\small (0,0,0,1)}}^{0}$ & $\otimes $ & $\mathbf{6}_{{\small %
(0,0,0,1)}}^{0}$ & $=$ & $\mathbf{36}_{{\small (0,0,0,1)}}^{0}$ \\ 
$\mathbf{6}_{{\small (0,0,0,1)}}^{0}$ & $\otimes $ & $\mathbf{5}_{{\small %
(1,-1,-1,0)}}^{+}$ & $=$ & $\mathbf{30}_{{\small (0,0,0,0)}}^{0}$ \\ 
$\mathbf{6}_{{\small (0,0,0,1)}}^{0}$ & $\otimes $ & $\mathbf{5}_{{\small %
(-1,-1,1,0)}}^{-}$ & $=$ & $\mathbf{30}_{{\small (0,0,0,0)}}^{0}$ \\ 
$\mathbf{6}_{{\small (0,0,0,1)}}^{0}$ & $\otimes $ & $\mathbf{4}_{{\small %
(2,1,0,-1)}}^{+}$ & $=$ & $\mathbf{24}_{{\small (0,0,0,-1)}}^{0}$ \\ 
$\mathbf{6}_{{\small (0,0,0,1)}}^{0}$ & $\otimes $ & $\mathbf{4}_{{\small %
(-2,1,0,-1)}}^{-}$ & $=$ & $\mathbf{24}_{{\small (0,0,0,-1)}}^{0}$ \\ 
$\mathbf{6}_{{\small (0,0,0,1)}}^{0}$ & $\otimes $ & $\mathbf{1}_{{\small %
(-1,1,-1,1)}}^{-}$ & $=$ & $\mathbf{6}_{{\small (0,0,0,1)}}^{0}$%
\end{tabular}
\label{x}
\end{equation}%
and%
\begin{equation}
\begin{tabular}{lllll}
$\mathbf{5}_{{\small (1,-1,-1,0)}}^{+}$ & $\otimes $ & $\mathbf{5}_{{\small %
(1,-1,-1,0)}}^{+}$ & $=$ & $\mathbf{25}_{{\small (1,1,1,0)}}^{+}$ \\ 
$\mathbf{5}_{{\small (1,-1,-1,0)}}^{+}$ & $\otimes $ & $\mathbf{5}_{{\small %
(-1,-1,1,0)}}^{-}$ & $=$ & $\mathbf{25}_{{\small (-1,1,-1,0)}}^{-}$ \\ 
$\mathbf{5}_{{\small (1,-1,-1,0)}}^{+}$ & $\otimes $ & $\mathbf{4}_{{\small %
(2,1,0,-1)}}^{+}$ & $=$ & $\mathbf{20}_{{\small (2,-1,0,0)}}^{+}$ \\ 
$\mathbf{5}_{{\small (1,-1,-1,0)}}^{+}$ & $\otimes $ & $\mathbf{4}_{{\small %
(-2,1,0,-1)}}^{-}$ & $=$ & $\mathbf{20}_{{\small (-2,-1,0,0)}}^{-}$ \\ 
$\mathbf{5}_{{\small (1,-1,-1,0)}}^{+}$ & $\otimes $ & $\mathbf{1}_{{\small %
(-1,1,-1,1)}}^{-}$ & $=$ & $\mathbf{5}_{{\small (-1,-1,1,0)}}^{-}$ \\ 
$\mathbf{5}_{{\small (-1,-1,1,0)}}^{-}$ & $\otimes $ & $\mathbf{5}_{{\small %
(-1,-1,1,0)}}^{-}$ & $=$ & $\mathbf{25}_{{\small (1,1,1,0)}}^{+}$ \\ 
$\mathbf{5}_{{\small (-1,-1,1,0)}}^{-}$ & $\otimes $ & $\mathbf{4}_{{\small %
(2,1,0,-1)}}^{+}$ & $=$ & $\mathbf{20}_{{\small (-2,-1,0,0)}}^{-}$ \\ 
$\mathbf{5}_{{\small (-1,-1,1,0)}}^{-}$ & $\otimes $ & $\mathbf{4}_{{\small %
(-2,1,0,-1)}}^{-}$ & $=$ & $\mathbf{20}_{{\small (2,-1,0,0)}}^{+}$ \\ 
$\mathbf{5}_{{\small (-1,-1,1,0)}}^{-}$ & $\otimes $ & $\mathbf{1}_{{\small %
(-1,1,-1,1)}}^{-}$ & $=$ & $\mathbf{5}_{{\small (1,-1,-1,0)}}^{+}$%
\end{tabular}
\label{y}
\end{equation}%
as well as%
\begin{equation}
\begin{tabular}{lllll}
$\mathbf{4}_{{\small (2,1,0,-1)}}^{+}$ & $\otimes $ & $\mathbf{4}_{{\small %
(2,1,0,-1)}}^{+}$ & $=$ & $\mathbf{16}_{{\small (4,1,0,1)}}^{+}$ \\ 
$\mathbf{4}_{{\small (2,1,0,-1)}}^{+}$ & $\otimes $ & $\mathbf{4}_{{\small %
(-2,1,0,-1)}}^{-}$ & $=$ & $\mathbf{16}_{{\small (-4,1,0,1)}}^{-}$ \\ 
$\mathbf{4}_{{\small (2,1,0,-1)}}^{+}$ & $\otimes $ & $\mathbf{1}_{{\small %
(-1,1,-1,1)}}^{-}$ & $=$ & $\mathbf{4}_{{\small (-2,1,0,-1)}}^{-}$ \\ 
$\mathbf{4}_{{\small (-2,1,0,-1)}}^{-}$ & $\otimes $ & $\mathbf{4}_{{\small %
(-2,1,0,-1)}}^{-}$ & $=$ & $\mathbf{16}_{{\small (4,1,0,1)}}^{+}$ \\ 
$\mathbf{4}_{{\small (-2,1,0,-1)}}^{-}$ & $\otimes $ & $\mathbf{1}_{{\small %
(-1,1,-1,1)}}^{-}$ & $=$ & $\mathbf{4}_{{\small (2,1,0,-1)}}^{+}$ \\ 
$\mathbf{1}_{{\small (-1,1,-1,1)}}^{-}$ & $\otimes $ & $\mathbf{1}_{{\small %
(-1,1,-1,1)}}^{-}$ & $=$ & $\mathbf{1}_{{\small (1,1,1,1)}}^{+}$%
\end{tabular}
\label{z}
\end{equation}%
with right sides obtained by requiring conservation of characters; we find:%
\begin{equation}
\begin{tabular}{lllll}
$\mathbf{36}_{{\small (0,0,0,1)}}^{0}$ & $=$ & $\mathbf{30}_{{\small %
(0,0,0,0)}}^{0}$ & $\oplus $ & $\mathbf{6}_{{\small (0,0,0,1)}}^{0}$ \\ 
$\mathbf{30}_{{\small (0,0,0,0)}}^{0}$ & $=$ & $\mathbf{24}_{{\small %
(0,0,0,-1)}}^{0}$ & $\oplus $ & $\mathbf{6}_{{\small (0,0,0,1)}}^{0}$ \\ 
$\mathbf{25}_{{\small (1,1,1,0)}}^{+}$ & $=$ & $\mathbf{24}_{{\small %
(0,0,0,-1)}}^{0}$ & $\oplus $ & $\boldsymbol{1}_{{\small (1,1,1,1)}}^{+}$ \\ 
$\mathbf{25}_{{\small (-1,1,-1,0)}}^{-}$ & $=$ & $\mathbf{24}_{{\small %
(0,0,0,-1)}}^{0}$ & $\oplus $ & $\boldsymbol{1}_{{\small (-1,1,-1,1)}}^{-}$
\\ 
$\mathbf{24}_{{\small (0,0,0,-1)}}^{0}$ & $=$ & $\mathbf{18}_{{\small %
(0,0,0,-2)}}^{0}$ & $\oplus $ & $\mathbf{6}_{{\small (0,0,0,1)}}^{0}$ \\ 
$\mathbf{20}_{{\small (2,-1,0,0)}}^{+}$ & $=$ & $\mathbf{14}_{{\small %
(2,-1,0,-1)}}^{+}$ & $\oplus $ & $\mathbf{6}_{{\small (0,0,0,1)}}^{0}$ \\ 
$\mathbf{20}_{{\small (-2,-1,0,0)}}^{-}$ & $=$ & $\mathbf{14}_{{\small %
(-2,-1,0,-1)}}^{-}$ & $\oplus $ & $\mathbf{6}_{{\small (0,0,0,1)}}^{0}$ \\ 
$\mathbf{16}_{{\small (4,1,0,1)}}^{+}$ & $=$ & $\mathbf{10}_{{\small %
(4,1,0,0)}}^{+}$ & $\oplus $ & $\mathbf{6}_{{\small (0,0,0,1)}}^{0}$ \\ 
$\mathbf{16}_{{\small (-4,1,0,1)}}^{-}$ & $=$ & $\mathbf{10}_{{\small %
(-4,1,0,0)}}^{-}$ & $\oplus $ & $\mathbf{6}_{{\small (0,0,0,1)}}^{0}$%
\end{tabular}
\label{t}
\end{equation}%
where we have set 
\begin{equation}
\begin{tabular}{lll}
$\mathbf{18}_{{\small (0,0,0,-2)}}^{0}$ & $=$ & $\mathbf{10}_{{\small %
(0,-2,0,0)}}^{0}\oplus \mathbf{4}_{{\small (2,1,0,-1)}}^{+}\oplus \mathbf{4}%
_{{\small (-2,1,0,-1)}}^{-}$ \\ 
$\mathbf{14}_{{\small (2,-1,0,-1)}}^{+}$ & $=$ & $\mathbf{10}_{{\small %
(0,-2,0,0)}}^{0}\oplus \mathbf{4}_{{\small (2,1,0,-1)}}^{+}$ \\ 
$\mathbf{14}_{{\small (-2,-1,0,-1)}}^{-}$ & $=$ & $\mathbf{10}_{{\small %
(0,-2,0,0)}}^{0}\oplus \mathbf{4}_{{\small (-2,1,0,-1)}}^{-}$%
\end{tabular}
\label{u}
\end{equation}%
and%
\begin{equation}
\begin{tabular}{lll}
$\mathbf{10}_{{\small (0,-2,0,0)}}^{0}$ & $=$ & $\mathbf{5}_{{\small %
(1,-1,-1,0)}}^{+}\oplus \mathbf{5}_{{\small (-1,-1,1,0)}}^{-}$ \\ 
$\mathbf{10}_{{\small (4,1,0,0)}}^{+}$ & $=$ & $\mathbf{5}_{{\small %
(1,-1,-1,0)}}^{+}\oplus \mathbf{4}_{{\small (2,1,0,-1)}}^{+}\oplus 
\boldsymbol{1}_{{\small (1,1,1,1)}}^{+}$ \\ 
$\mathbf{10}_{{\small (-4,1,0,0)}}^{-}$ & $=$ & $\mathbf{5}_{{\small %
(-1,-1,1,0)}}^{-}\oplus \mathbf{4}_{{\small (-2,1,0,-1)}}^{-}\oplus 
\boldsymbol{1}_{{\small (-1,1,-1,1)}}^{-}$%
\end{tabular}
\label{v}
\end{equation}%
Putting eqs(\ref{t}-\ref{v}) back into eqs(\ref{x}-\ref{z}), one obtains the
full explicit expression of the fusion algebra $\mathcal{F}_{\mathbb{S}_{5}}$%
.

\subsection{Fusion algebra $\mathcal{F}_{\mathbb{A}_{5}}$}

The alternating $\mathbb{A}_{5}$ is an order \emph{60} subgroup of the
symmetric $\mathbb{S}_{5}$; it has \emph{5} conjugacy classes $\mathfrak{C}%
_{i}$ and \emph{5} irreducible representations $\boldsymbol{R}_{i}$ with
dimensions as in the expansion 
\begin{equation}
60=1^{2}+3^{2}+\left( 3^{\prime }\right) ^{2}+4^{2}+5^{2}
\end{equation}%
The $\mathbb{A}_{5}$ group has \emph{3} non commuting generators $\alpha ,$ $%
\beta ,$ $\gamma $, obeying $\alpha ^{2}=\beta ^{3}=\gamma ^{5}=\alpha \beta
\gamma =1$, with characters given by real numbers as follows; see also table
(\ref{3}) in appendix,%
\begin{equation}
\begin{tabular}{|l|l|l|l|l|l|}
\hline
${\small \chi }_{_{\boldsymbol{R}_{i}}}^{g}$ & ${\small \  \  \chi }_{_{%
\mathbf{1}}}$ & ${\small \  \  \chi }_{_{\mathbf{3}}}$ & ${\small \  \  \chi }%
_{_{\mathbf{3}^{\prime }}}$ & ${\small \  \  \chi }_{\mathbf{4}}$ & ${\small \
\  \chi }_{_{5}}$ \\ \hline
${\small \alpha }$ & ${\small \  \ 1}$ & ${\small -1}$ & ${\small -1}$ & $%
{\small \  \ 0}$ & ${\small \  \ 1}$ \\ \hline
${\small \beta }$ & ${\small \  \ 1}$ & ${\small \  \ 0}$ & ${\small \  \ 0}$ & 
${\small \  \ 1}$ & ${\small -1}$ \\ \hline
${\small \gamma }$ & ${\small \  \ 1}$ & ${\small \  \  \kappa }_{+}$ & $%
{\small \  \  \kappa }_{-}$ & ${\small -1}$ & ${\small \  \ 0}$ \\ \hline
\end{tabular}%
\end{equation}%
where $\kappa _{\pm }=\frac{1\pm \sqrt{5}}{2}$. We denote the \emph{5}
irreducible representations of $\mathbb{A}_{5}$ like 
\begin{equation}
\begin{tabular}{lllll}
$\mathbf{1}=\boldsymbol{1}_{{\small (1,1,1)}}^{0}$ & , & $\mathbf{3}=%
\boldsymbol{3}_{{\small (-1,0,\kappa }_{+}{\small )}}^{+}$ & , & $\mathbf{3}=%
\boldsymbol{3}_{{\small (-1,0,\kappa }_{-}{\small )}}^{-}$ \\ 
$\mathbf{4}=\boldsymbol{4}_{{\small (0,1,-1)}}^{0}$ & , & $\mathbf{5}=%
\boldsymbol{5}_{{\small (1,-1,0)}}^{0}$ &  & 
\end{tabular}%
\end{equation}%
The obtained fusion algebra $\mathcal{F}_{\mathbb{A}_{5}}$ of the $\mathbb{A}%
_{5}$- irreducible representations is given by%
\begin{equation}
\begin{tabular}{lllll}
$\mathbf{5}_{{\small (1,-1,0)}}^{0}$ & $\otimes $ & $\mathbf{5}_{{\small %
(1,-1,0)}}^{0}$ & $=$ & $\mathbf{25}_{{\small (1,1,0)}}^{0}$ \\ 
$\mathbf{5}_{{\small (1,-1,0)}}^{0}$ & $\otimes $ & $\mathbf{4}_{{\small %
(0,1,-1)}}^{0}$ & $=$ & $\mathbf{20}_{{\small (0,-1,0)}}^{0}$ \\ 
$\mathbf{5}_{{\small (1,-1,0)}}^{0}$ & $\otimes $ & $\mathbf{3}_{{\small %
(-1,0,\kappa }_{+}{\small )}}^{+}$ & $=$ & $\mathbf{15}_{{\small (-1,0,0)}%
}^{0}$ \\ 
$\mathbf{5}_{{\small (1,-1,0)}}^{0}$ & $\otimes $ & $\mathbf{3}_{{\small %
(-1,0,\kappa }_{-}{\small )}}^{-}$ & $=$ & $\mathbf{15}_{{\small (-1,0,0)}%
}^{0}$%
\end{tabular}%
\end{equation}%
and%
\begin{equation}
\begin{tabular}{lllll}
$\mathbf{4}_{{\small (0,1,-1)}}^{0}$ & $\otimes $ & $\mathbf{4}_{{\small %
(0,1,-1)}}^{0}$ & $=$ & $\mathbf{16}_{{\small (0,1,-1)}}^{0}$ \\ 
$\mathbf{4}_{{\small (0,1,-1)}}^{0}$ & $\otimes $ & $\mathbf{3}_{{\small %
(-1,0,\kappa }_{+}{\small )}}^{+}$ & $=$ & $\mathbf{12}_{{\small %
(0,0,-\kappa }_{+}{\small )}}^{+}$ \\ 
$\mathbf{4}_{{\small (0,1,-1)}}^{0}$ & $\otimes $ & $\mathbf{3}_{{\small %
(-1,0,\kappa }_{-}{\small )}}^{-}$ & $=$ & $\mathbf{12}_{{\small %
(0,0,-\kappa }_{-}{\small )}}^{-}$%
\end{tabular}
\label{m}
\end{equation}%
as well as%
\begin{equation}
\begin{tabular}{lllll}
$\mathbf{3}_{{\small (-1,0,\kappa }_{+}{\small )}}^{+}$ & $\otimes $ & $%
\mathbf{3}_{{\small (-1,0,\kappa }_{+}{\small )}}^{+}$ & $=$ & $\mathbf{9}_{%
{\small (1,0,\kappa }_{+}^{2}{\small )}}^{+}$ \\ 
$\mathbf{3}_{{\small (-1,0,\kappa }_{+}{\small )}}^{+}$ & $\otimes $ & $%
\mathbf{3}_{{\small (-1,0,\kappa }_{-}{\small )}}^{-}$ & $=$ & $\mathbf{9}_{%
{\small (1,0,\kappa _{+}\kappa }_{-}{\small )}}^{0}$ \\ 
$\mathbf{3}_{{\small (-1,0,\kappa }_{-}{\small )}}^{-}$ & $\otimes $ & $%
\mathbf{3}_{{\small (-1,0,\kappa }_{-}{\small )}}^{-}$ & $=$ & $\mathbf{9}_{%
{\small (1,0,\kappa }_{-}^{2}{\small )}}^{-}$%
\end{tabular}
\label{n}
\end{equation}%
where right hand sides of eqs(\ref{m}-\ref{n}) are given by 
\begin{equation}
\begin{tabular}{lllll}
$\mathbf{25}_{{\small (1,1,0)}}^{0}$ & $=$ & $\mathbf{20}_{{\small (0,-1,0)}%
}^{0}$ & $\oplus $ & $\mathbf{4}_{{\small (0,1,-1)}}^{0}\oplus \boldsymbol{1}%
_{{\small (1,1,1)}}^{0}$ \\ 
$\mathbf{20}_{{\small (0,-1,0)}}^{0}$ & $=$ & $\mathbf{15}_{{\small (-1,0,0)}%
}^{0}$ & $\oplus $ & $\mathbf{5}_{{\small (1,-1,0)}}^{0}$ \\ 
$\mathbf{16}_{{\small (0,1,1)}}$ & $=$ & $\mathbf{15}_{{\small (-1,0,0)}%
}^{0} $ & $\oplus $ & $\boldsymbol{1}_{{\small (1,1,1)}}^{0}$ \\ 
$\mathbf{15}_{{\small (-1,0,0)}}^{0}$ & $=$ & $\mathbf{9}_{{\small (1,0,-1)}%
}^{0}$ & $\oplus $ & $\mathbf{6}_{{\small (-2,0,1)}}^{0}$ \\ 
$\mathbf{6}_{{\small (-2,0,1)}}^{0}$ & $=$ & $\mathbf{3}_{{\small %
(-1,0,\kappa }_{+}{\small )}}^{+}$ & $\oplus $ & $\mathbf{3}_{{\small %
(-1,0,\kappa }_{-}{\small )}}^{-}$%
\end{tabular}%
\end{equation}%
and%
\begin{equation}
\begin{tabular}{lllll}
$\mathbf{12}_{{\small (0,0,-\kappa }_{+}{\small )}}^{+}$ & $=$ & $\mathbf{9}%
_{{\small (1,0,-1)}}^{0}$ & $\oplus $ & $\mathbf{3}_{{\small (-1,0,\kappa }%
_{-}{\small )}}^{-}$ \\ 
$\mathbf{12}_{{\small (0,0,-\kappa }_{-}{\small )}}^{-}$ & $=$ & $\mathbf{9}%
_{{\small (1,0,-1)}}^{0}$ & $\oplus $ & $\mathbf{3}_{{\small (-1,0,\kappa }%
_{+}{\small )}}^{+}$ \\ 
$\mathbf{9}_{{\small (1,0,-1)}}^{0}$ & $=$ & $\mathbf{5}_{{\small (1,-1,0)}%
}^{0}$ & $\oplus $ & $\mathbf{4}_{{\small (0,1,-1)}}^{0}$ \\ 
$\mathbf{9}_{{\small (1,0,\kappa }_{+}^{2}{\small )}}^{+}$ & $=$ & $\mathbf{5%
}_{{\small (1,-1,0)}}^{0}$ & $\oplus $ & $\mathbf{3}_{{\small (-1,0,\kappa }%
_{+}{\small )}}^{+}\oplus \boldsymbol{1}_{{\small (1,1,1)}}^{0}$ \\ 
$\mathbf{9}_{{\small (1,0,\kappa _{-}^{2})}}^{-}$ & $=$ & $\mathbf{5}_{%
{\small (1,-1,0)}}^{0}$ & $\oplus $ & $\mathbf{3}_{{\small (-1,0,\kappa }_{-}%
{\small )}}^{-}\oplus \boldsymbol{1}_{{\small (1,1,1)}}^{0}$%
\end{tabular}%
\end{equation}

\section{Conclusion}

F-theory GUT models have a finite spectrum of localised matter curves $%
\left
\{ \Phi _{R_{i}}\right \} $ indexed, in addition to gauge charges, by
quantum numbers of monodromy $\Gamma $.\ In the example of $SO_{10}\times
\Gamma $ models, the possible $\Gamma $'s are given by subgroups of the
symmetric $\mathbb{S}_{4}$ as depicted on figure \ref{5}. In the particular $%
SO_{10}\times \mathbb{S}_{4}$ model, localised matter curves is as in table (%
\ref{a3}) involving, in addition to 16-plets and 10-plets, flavons $%
\vartheta $ carrying non trivial charges under monodromy $\mathbb{S}_{4}$.
The same properties hold for $SO_{10}\times \Gamma $ models with monodromy $%
\Gamma \subset \mathbb{S}_{4}$; the main difference is that now the numbers $%
n_{i}^{\Gamma }$\ and the dimensions $d_{i}^{\Gamma }$ of irreducible
representations $\boldsymbol{R}_{i}^{\Gamma }$ of monodromy group $\Gamma $\
are smaller as shown on eqs(\ref{41},\ref{42},\ref{43}); but the
corresponding matter spectrums $\left \{ \Phi _{R_{i}}\right \} $ still
carry charges under $\Gamma $ including the flavons which play an important
role in models building; in particular in the study of neutrino mixing and
the Higgs sector of extended MSSM as well as for dealing with GUT
constraints such as proton decay\textrm{. }By requiring invariance under
monodromy $\Gamma $, one disposes therefore of an important tool to deal
with constructing general superpotentials involving flavons. This
construction, requires however the fusion rules (\ref{j}). The same think
can be said about $SU_{5}\times \mathbb{S}_{5}$ prototype and in general
about the $SU_{5}\times \Gamma $ models where monodromies $\Gamma $ are
given by subgroups of $\mathbb{S}_{5}$.\newline
In this work we have constructed the closed fusion algebras $\mathcal{F}%
_{\Gamma }$ of the F-GUT operators spectrum $\left \{ \Phi _{R_{i}}\right \} 
$ indexed by $\boldsymbol{R}_{i}^{\Gamma }$ representations of monodromy
groups $\Gamma $; in particular monodromy given by the symmetric groups $%
\mathbb{S}_{5},$ $\mathbb{S}_{4}$, $\mathbb{S}_{3};$ the non abelian
alternating $\mathbb{A}_{5}$, $\mathbb{A}_{4}$; and the dihedral $\mathbb{D}%
_{4}$. These $\mathcal{F}_{\Gamma }$'s are important for building monodromy
invariant superpotentials $W\left( \Phi _{R_{i}}\right) $ for GUT models
with gauge symmetry $G=SO_{10}$ and $SU_{5}$. In the example of $%
SO_{10}\times \Gamma $ theory, typical mass terms, generated by restricting
to VEVs of Higgs and flavons, have the form $16^{i}M_{ij}16^{j}$ with mass
matrix controlled by $\mathcal{F}_{\Gamma }$ fusion relations.\newline
To derive the $\mathcal{F}_{\Gamma }$'s structures obtained in this paper,
we have used properties of the characters of the irreducible representations
of discrete symmetries. Our construction, which may be used for other
purposes, extends straightforwardly to any finite symmetry group; including
products like $\Gamma _{1}\times \Gamma _{2}$. We end this study by
describing briefly the case of abelian symmetries $\boldsymbol{H}$; they
have completely reducible representations; and so simpler fusion algebras $%
\mathcal{F}_{\boldsymbol{H}}$. In the example of $\mathbb{Z}_{2}$, there are
two irreducible representations $\mathbf{1}_{\pm }$; and the corresponding
fusion algebra $\mathcal{F}_{\mathbb{Z}_{2}}$ is just $\mathbf{1}_{+}\otimes 
\mathbf{1}_{+}=\mathbf{1}_{+},$ $\mathbf{1}_{-}\otimes \mathbf{1}_{-}=%
\mathbf{1}_{+}$ and $\mathbf{1}_{-}\otimes \mathbf{1}_{+}=\mathbf{1}_{-}$.
For the case $\mathbb{Z}_{3}$, we have three irreducible 1-$\dim $
representations following from the $\mathbb{Z}_{3}$ group property 
\begin{equation*}
3=1^{2}+\left( 1^{\prime }\right) ^{2}+\left( 1^{\prime \prime }\right) ^{2}
\end{equation*}%
with characters given by the three cubic roots $j^{p}$ of unity; $j^{3}=1$.
By denoting these representations as $\mathbf{1}_{j^{p}}$ with $p=0,1,2$ $%
\func{mod}$ $3$, the corresponding fusion algebra $\mathcal{F}_{\mathbb{Z}%
_{3}}$ is nothing but $\mathbf{1}_{j^{p}}\otimes \mathbf{1}_{j^{q}}=\mathbf{1%
}_{j^{p+q}}$. Extension to $\mathcal{F}_{\mathbb{Z}_{N}}$ is straightforward.

\begin{acknowledgement}
I thank the International Centre of Theoretical Physics, ICTP, Trieste-
Italy, where this work has been done.
\end{acknowledgement}

\section{Appendix}

In this appendix, we collect the character tables of discrete $\Gamma $'s
appearing in F-theory GUT. The $\mathfrak{C}_{i}$'s refer to conjugacy
classes; the $\boldsymbol{R}_{i}$'s for irreducible representations, and $%
\mathrm{\chi }_{_{\boldsymbol{R}_{i}}}$'s to characters.

\begin{itemize}
\item \emph{permutation symmetry} $\mathbb{S}_{4}$%
\begin{equation}
\begin{tabular}{|l|l|l|l|l|l|l|}
\hline
$\mathfrak{C}_{i}$\TEXTsymbol{\backslash}irrep $\boldsymbol{R}_{j}$ & $\  \ 
\mathrm{\chi }_{_{\boldsymbol{I}}}$ & $\  \  \mathrm{\chi }_{_{\boldsymbol{3}%
^{\prime }}}$ & $\  \  \mathrm{\chi }_{_{\boldsymbol{2}}}$ & $\  \  \mathrm{\chi 
}_{_{\boldsymbol{3}}}$ & $\  \  \mathrm{\chi }_{_{\epsilon }}$ & {\small order}
\\ \hline
$\mathfrak{C}_{1}\equiv \mathrm{\  \ e}$ & $\  \ 1\  \ $ & $\  \ 3\  \ $ & $\  \
2\  \ $ & $\  \ 3\  \ $ & $\  \ 1\  \ $ & $\  \ 1\  \ $ \\ \hline
$\mathfrak{C}_{2}\equiv \mathrm{(12)}$ & $\  \ 1$ & $-1$ & $\  \ 0$ & $\  \ 1$
& $-1$ & $\  \ 6$ \\ \hline
$\mathfrak{C}_{3}\equiv \mathrm{(12)(34)}$ & $\  \ 1$ & $-1$ & $\  \ 2$ & $-1$
& $\  \ 1$ & $\  \ 3$ \\ \hline
$\mathfrak{C}_{4}\equiv \mathrm{(123)}$ & $\  \ 1$ & $\  \ 0$ & $-1$ & $\  \ 0$
& $\  \ 1$ & $\  \ 8$ \\ \hline
$\mathfrak{C}_{5}\equiv \mathrm{(1234)}$ & $\  \ 1$ & $\  \ 1$ & $\  \ 0$ & $-1$
& $-1$ & $\  \ 6$ \\ \hline
\end{tabular}
\label{1}
\end{equation}

\item \emph{permutation symmetry} $\mathbb{S}_{3}$%
\begin{equation}
\begin{tabular}{|l|l|l|l|l|}
\hline
$\mathfrak{C}_{i}$\TEXTsymbol{\backslash}irrep $\boldsymbol{R}_{j}$ & $%
\mathbf{\chi }_{_{\boldsymbol{I}}}$ & $\mathbf{\chi }_{_{\boldsymbol{2}}}$ & 
$\mathbf{\chi }_{_{\epsilon }}$ & {\small order} \\ \hline
$\mathfrak{C}_{1}\equiv \mathrm{\  \ e}$ & $\  \ 1$ \  & $\  \ 2$ \  & $\  \ 1$
\  & $\  \ 1$ \\ \hline
$\mathfrak{C}_{2}\equiv \mathrm{(12)}$ & $\  \ 1$ & $\ 0$ & $-1$ & $\  \ 3$ \\ 
\hline
$\mathfrak{C}_{3}\equiv \mathrm{(123)}$ & $\  \ 1$ & $-1$ & $\  \ 1$ & $\  \ 2$
\\ \hline
\end{tabular}
\label{33}
\end{equation}

\item \emph{alternating group }$\mathbb{A}_{4}$%
\begin{equation}
\begin{tabular}{|l|l|l|l|l|l|}
\hline
$\mathfrak{C}_{i}$\TEXTsymbol{\backslash}irrep $\boldsymbol{R}_{j}$ & $\  \ 
\mathrm{\chi }_{_{\boldsymbol{I}}}$ & $\  \  \mathrm{\chi }_{_{\boldsymbol{1}%
^{\prime }}}$ & $\  \  \mathrm{\chi }_{_{\boldsymbol{1}^{\prime \prime }}}$ & $%
\  \  \mathrm{\chi }_{_{\boldsymbol{3}}}$ & {\small order} \\ \hline
$\mathfrak{C}_{1}\equiv \mathrm{\  \ e}$ & $\  \ 1\  \ $ & $\  \ 1\  \ $ & $\  \
1\  \ $ & $\  \ 3\  \ $ & $\  \ 1\  \ $ \\ \hline
$\mathfrak{C}_{2}\equiv \mathrm{(12)(34)}$ & $\  \ 1$ & $\  \ 1$ & $\  \ 1$ & $%
-1$ & $\  \ 3$ \\ \hline
$\mathfrak{C}_{3}\equiv \mathrm{(123)}$ & $\  \ 1$ & $\  \ j$ & $\  \ j^{2}$ & $%
\  \ 0$ & $\  \ 4$ \\ \hline
$\mathfrak{C}_{4}\equiv \mathrm{(132)}$ & $\  \ 1$ & $\  \ j^{2}$ & $\  \ j$ & $%
\ 0$ & $\  \ 4$ \\ \hline
\end{tabular}
\label{22}
\end{equation}

\item \emph{dihedral symmetry} $\mathbb{D}_{4}$\newline
\begin{equation}
\begin{tabular}{|l|l|l|l|l|l|l|}
\hline
$\mathfrak{C}_{i}$\TEXTsymbol{\backslash}$\mathrm{\chi }_{\boldsymbol{R}%
_{j}} $ & $\  \  \mathrm{\chi }_{_{\mathbf{1}_{1}}}$ & $\  \  \mathrm{\chi }_{_{%
\mathbf{1}_{2}}}$ & $\  \  \mathrm{\chi }_{_{\mathbf{1}_{3}}}$ & $\  \  \mathrm{%
\chi }_{\mathbf{1}_{4}}$ & $\  \  \mathrm{\chi }_{_{2}}$ & {\small order} \\ 
\hline
$\mathfrak{C}_{1}$ & $\  \ 1\  \ $ & $\  \ 1\  \ $ & $\  \ 1\  \ $ & $\  \ 1\  \ $ & 
$\  \ 2\  \ $ & $\  \ 1\  \ $ \\ \hline
$\mathfrak{C}_{2}$ & $\  \ 1$ & $\  \ 1$ & $\  \ 1$ & $\  \ 1$ & $-2$ & $\  \ 1$
\\ \hline
$\mathfrak{C}_{3}$ & $\  \ 1$ & $\  \ 1$ & $-1$ & $-1$ & $\  \ 0$ & $\  \ 2$ \\ 
\hline
$\mathfrak{C}_{4}$ & $\  \ 1$ & $-1$ & $\  \ 1$ & $-1$ & $\  \ 0$ & $\  \ 2$ \\ 
\hline
$\mathfrak{C}_{5}$ & $\  \ 1$ & $-1$ & $-1$ & $\  \ 1$ & $-0$ & $\  \ 2$ \\ 
\hline
\end{tabular}
\label{hc}
\end{equation}

\item \emph{permutation symmetry} $\mathbb{S}_{5}$%
\begin{equation}
\begin{tabular}{|l|l|l|l|l|l|l|l|l|}
\hline
$\mathfrak{C}_{i}$\TEXTsymbol{\backslash}$\mathrm{\chi }_{\boldsymbol{R}%
_{j}} $ & $\mathbf{\  \  \chi }_{_{\mathbf{1}}}$ & $\mathbf{\  \  \chi }_{_{%
\mathbf{1}^{\prime }}}$ & $\mathbf{\  \  \chi }_{_{\mathbf{4}}}$ & $\mathbf{\
\  \chi }_{\mathbf{4}^{\prime }}$ & $\mathbf{\  \  \chi }_{_{5}}$ & $\mathbf{\
\  \chi }_{_{5^{\prime }}}$ & $\mathbf{\  \  \chi }_{_{6}}$ & {\small order} \\ 
\hline
$\mathfrak{C}_{1}=e$ & $\  \ 1\  \ $ & $\  \ 1\  \ $ & $\  \ 4\  \ $ & $\  \ 4\  \ $
& $\  \ 5\  \ $ & $\  \ 5$ & $\ 6$ & $\  \ 1\  \ $ \\ \hline
$\mathfrak{C}_{2}=\left( 12\right) $ & $\  \ 1$ & $-1$ & $\  \ 2$ & $-2$ & $\
1 $ & $-1$ & $\ 0$ & $\  \ 10$ \\ \hline
$\mathfrak{C}_{3}=\left( 12\right) \left( 34\right) $ & $\  \ 1$ & $\  \ 1$ & $%
\  \ 0$ & $\  \ 0$ & $\  \ 1$ & $\ 1$ & $-2$ & $\  \ 15$ \\ \hline
$\mathfrak{C}_{4}=\left( 123\right) $ & $\  \ 1$ & $\  \ 1$ & $\  \ 1$ & $\  \ 1$
& $-1$ & $-1$ & $\ 0$ & $\  \ 20$ \\ \hline
$\mathfrak{C}_{5}=\left( 1234\right) $ & $\  \ 1$ & $-1$ & $\ 0$ & $\  \ 0$ & $%
-1$ & $\ 1$ & $\ 0$ & $\  \ 30$ \\ \hline
$\mathfrak{C}_{6}=\left( 123\right) \left( 45\right) $ & $\  \ 1$ & $-1$ & $%
-1 $ & $\ 1$ & $\  \ 1$ & $-1$ & $\ 0$ & $\  \ 20$ \\ \hline
$\mathfrak{C}_{7}=\left( 12345\right) $ & $\  \ 1$ & $\ 1$ & $-1$ & $-1$ & $\
0$ & $\ 0$ & $\ 1$ & $\  \ 24$ \\ \hline
\end{tabular}
\label{2}
\end{equation}

\item \emph{alternating group }$\mathbb{A}_{5}$%
\begin{equation}
\begin{tabular}{|l|l|l|l|l|l|l|}
\hline
$\mathfrak{C}_{i}$\TEXTsymbol{\backslash}$\mathrm{\chi }_{\boldsymbol{R}%
_{j}} $ & $\mathbf{\  \  \chi }_{_{\mathbf{1}}}$ & $\mathbf{\  \  \chi }_{_{%
\mathbf{3}}}$ & $\mathbf{\  \  \chi }_{_{\mathbf{3}^{\prime }}}$ & $\mathbf{\
\  \chi }_{\mathbf{4}}$ & $\mathbf{\  \  \chi }_{_{5}}$ & {\small order} \\ 
\hline
$\mathfrak{C}_{1}=e$ & $\  \ 1\  \ $ & $\  \ 3\  \ $ & $\  \ 3\  \ $ & $\  \ 4\  \ $
& $\  \ 5\  \ $ & $\  \ 1\  \ $ \\ \hline
$\mathfrak{C}_{2}=\left( 12\right) \left( 34\right) $ & $\  \ 1$ & $-1$ & $-1$
& $\  \ 0$ & $\  \ 1$ & $\  \ 15$ \\ \hline
$\mathfrak{C}_{3}=\left( 123\right) $ & $\  \ 1$ & $\  \ 0$ & $\  \ 0$ & $\  \ 1$
& $-1$ & $\  \ 20$ \\ \hline
$\mathfrak{C}_{4}=\left( 12345\right) $ & $\  \ 1$ & $\  \  \kappa _{+}$ & $\  \
\kappa _{-}$ & $-1$ & $\  \ 0$ & $\  \ 12$ \\ \hline
$\mathfrak{C}_{5}=\left( 13524\right) $ & $\  \ 1$ & $\  \  \kappa _{-}$ & $\  \
\kappa _{+}$ & $-1$ & $\ 0$ & $\  \ 12$ \\ \hline
\end{tabular}
\label{3}
\end{equation}%
\begin{equation*}
\end{equation*}%
with $\kappa _{\pm }=\frac{1\pm \sqrt{5}}{2},$ $\kappa _{+}+\kappa _{-}=1,$ $%
\kappa _{-}\kappa _{+}=-1,$ $\kappa _{+}^{2}=1+\kappa _{+}$ and $\kappa
_{-}^{2}=1+\kappa _{-}=2-\kappa _{+}$.%
\begin{equation*}
\end{equation*}
\end{itemize}

\end{document}